\documentclass[a4paper,11pt]{article}

\usepackage{jheppub} 

\usepackage[T1]{fontenc} 
\usepackage{empheq,mathtools}
\usepackage{cancel,slashed}
\usepackage{braket}

\def\arXiv#1{\href{http://arxiv.org/abs/#1}{arXiv:#1}}
\def\arXiv#1#2{\href{http://arxiv.org/abs/#1}{arXiv:#1}}
\def\arXivid#1#2{\href{http://arxiv.org/abs/#1/#2}{#1/#2}}
\def\doi#1#2{\href{http://doi.org/#1}{#2}}

\title{\boldmath Holographic R\'{e}nyi entropies from hyperbolic black holes with scalar hair}

\author{Xiaoxuan Bai and Jie Ren}
\affiliation{School of Physics, Sun Yat-sen University, Guangzhou, 510275, China}

\emailAdd{baixx3@mail2.sysu.edu.cn}
\emailAdd{renjie7@mail.sysu.edu.cn}

\abstract{The R\'{e}nyi entropies as a generalization of the entanglement entropy imply much more information. We analytically calculate the R\'{e}nyi entropies (with a spherical entangling surface) by means of a class of neutral hyperbolic black holes with scalar hair as a one-parameter generalization of the MTZ black hole. The zeroth-order and third-order phase transitions of black holes lead to discontinuity of the R\'{e}nyi entropies and their second derivatives, respectively. From the R\'{e}nyi entropies that are analytic at $n=\infty$, we can express the entanglement spectrum as an infinite sum in terms of the Bell polynomials. We show that the analytic treatment is in agreement with numerical calculations for the low-lying entanglement spectrum in a wide range of parameters.}

\arxivnumber{2210.03732}

\begin{document} 
\maketitle
\flushbottom

\section{Introduction}
\label{sec:intro}
Entanglement entropy, which measures the entanglement of different parts in a quantum system, has considerable applications in condensed matter physics, e.g., the detection of topological ordered phases \cite{Levin:2006lw,Kitaev:2005dm}, quantum phase transitions \cite{Vidal:2002rm} and fractional quantum Hall effects \cite{Haldane:2008en} (see \cite{Amico:2007ag,Laflorencie:2015eck} for reviews). Moreover, the holographic description of the entanglement entropy \cite{Ryu:2006bv, Ryu:2006ef} in terms of the AdS/CFT correspondence may inspire a deeper understanding of the quantum gravity \cite{VanRaamsdonk:2010pw}. As a generalization of the entanglement entropy, the R\'{e}nyi entropies \cite{Renyi:1961me} imply much more information, e.g., with the R\'{e}nyi entropies of all orders, the entanglement spectrum \cite{Haldane:2008en} (eigenvalue distribution of the reduced density matrix) can be determined \cite{Calabrese:2008cl,Calabrese:2009qy,Hung:2011nu,Belin:2013dva,Guo:2020roc}.

A quantum system can be described by a density operator $\rho$. For simplicity, we consider a pure state consisting of two subsystems $A$ and $B$, characterized by vectors in Hilbert spaces $H_A$ and $H_B$, respectively. It is convenient to define the reduced density operator $\rho_A=\text{Tr}_B \, \rho_{AB}$, which only depends on the degrees of freedom of $A$. The entanglement (von Neumann) entropy is $S_{EE}=-\text{Tr}\rho_A \log\rho_A$, which becomes zero when $\rho_A$ describes a pure state, or equivalently, the whole system is characterized by $\rho_{AB}=\rho_A\otimes\rho_B$. With $n$ as a parameter, the R\'{e}nyi entropies are defined as
\begin{equation}
  S_{n}=\frac{1}{1-n}\log \text{Tr} [\rho_{A}^n],
\end{equation}
which return to the entanglement entropy in the $n\rightarrow 1$ limit.

Generally, these entropies can be calculated with the replica trick \cite{Holzhey:1994we,Calabrese:2004eu,Calabrese:2009qy} in quantum field theory by considering the Euclidean functional integral on $n$ copies of Riemann surfaces that are matched together. In terms of the AdS/CFT correspondence, the entanglement entropy can be obtained from general relativity by the Ryu-Takayanagi (RT) formula \cite{Ryu:2006bv, Ryu:2006ef}. In particular, for two regions separated by a spherical entangling surface, the R\'{e}nyi entropies of the ground state can be calculated by means of a hyperbolic (topological) black hole \cite{Casini:2011kv, Hung:2011nu} (we briefly review this approach in section~\ref{sec:ren-bh}). The temperature of the black hole is related to the R\'{e}nyi parameter $n$, where a larger $n$ corresponds to a lower temperature. 

The hyperbolic Schwarzschild-AdS (SAdS) solution is 
\begin{equation}
  ds^2=-\biggl(-1-\frac{2M}{r^{d-2}}+\frac{r^2}{L^2}\biggr)dt^2+\biggl(-1-\frac{2M}{r^{d-2}}+\frac{r^2}{L^2}\biggr)^{-1}dr^2+r^2d\Sigma_{d-1}^2,
  \label{eq:Sch-AdS}
\end{equation}
where $d\Sigma_{d-1}^2=du^2+\sinh^2{u}\, d\Omega_{d-2}^2$ is a $(d-1)$-dimensional hyperbolic space of unit radius. The hyperbolic SAdS black hole is stable at any temperature \cite{Emparan:1998he,Birmingham:1998nr}. However, if a scalar field is introduced to the system, the black hole may experience a phase transition to the one with scalar hair at a critical temperature. This can be implied from the fact that the IR geometry of the extremal hyperbolic black hole~\eqref{eq:Sch-AdS} includes an AdS$_2$ factor, which leads to the Breitenlohner-Freedman bound violated \cite{Dias:2010ma}. Accordingly, the R\'{e}nyi entropies also have a phase transition at a special $n$, which has been discussed in \cite{Belin:2013dva, Fang:2016ehk} in terms of numerical solutions of hairy hyperbolic black holes. In this paper, we provide more analytic calculations. 

The Einstein-Maxwell-dilaton (EMD) systems, with gauge field and dilaton field coupled to gravity, have wide applications in gauge/gravity duality. An especially useful class of analytic solutions of the EMD systems was found in \cite{Gao:2004tu}, with special cases belonging to gauged supergravity. In a particular neutral limit, the gauge field vanishes while the dilaton field is kept, resulting in neutral hyperbolic black holes with scalar hair \cite{Ren:2019lgw}. These black holes have zeroth-order and third-order phase transitions, and the corresponding R\'{e}nyi entropies and their second derivatives show discontinuity, respectively.

With the analytic solutions of the Einstein-scalar systems in hand, we explicitly obtain the R\'{e}nyi entropies $S_n$ and calculate the entanglement spectrum. For the one-parameter family of our systems, these R\'{e}nyi entropies converge to the same entanglement entropy in the $n\rightarrow 1$ limit. For special cases, we find that the R\'{e}nyi entropies have the same $(1+1/n)$ factor as the universal result in 2D CFTs. Starting with the analytic R\'{e}nyi entropies, we can calculate the entanglement spectrum by an inverse Laplace transform. Consequently, we can express the entanglement spectrum in terms of an infinite sum. By truncating this sum, we can verify that it is consistent with the numerical calculation for a wide range of eigenvalues.

The paper is organized as follows. In section~\ref{sec:ads4-renyi}, we begin by introducing a class of neutral hyperbolic black holes with scalar hair \cite{Ren:2019lgw} in AdS$_4$ spacetime and refining their thermodynamics and stability analysis. Then we calculate holographic R\'{e}nyi entropies and discuss corresponding phase transitions. Besides, we verify some R\'{e}nyi entropy inequalities in terms of the thermodynamically preferred solution. In section~\ref{sec:ads5-renyi}, we generalize these calculations to higher dimensions. In section~\ref{sec:spectrum}, we calculate the entanglement spectrum from the R\'{e}nyi entropies as an infinite sum in terms of the Bell polynomials and discuss its validity by comparing it with additional numerical calculations. In appendix~\ref{sec:bh-thermo}, we give the boundary counter terms for holographic renormalization, from which we obtain the mass and free energy of the hyperbolic black holes. In appendix~\ref{sec:math-notes}, we give a brief note on the mathematics related to our calculations, including the inverse Laplace transform and the exponential Bell polynomials.

\subsection{R\'{e}nyi entropies from hyperbolic black holes}
\label{sec:ren-bh}
For the case with a spherical entangling surface, we can calculate the R\'{e}nyi entropies by conformally mapping the causal development of the enclosed region to a hyperbolic cylinder $\mathbb{R} \times \mathbb{H}^{d-1}$ (see \cite{Casini:2011kv,Hung:2011nu} for details). Assuming that the radius of the entangling sphere is $L$, then the corresponding hyperbolic space $\mathbb{H}^{d-1}$ also has the radius $L$. Under this conformal transformation, the reduced density matrix is acted by a unitary operator
\begin{equation}
  \rho=U\frac{e^{-H/T_0}}{Z(T_0)}U^{-1},
\end{equation}
where $H$ is the modular Hamiltonian generating the (Rindler) time translation on $\mathbb{R} \times \mathbb{H}^{d-1}$, and $Z(T_0)=\text{Tr}[e^{H/T_0}]$ is the thermal partition function at temperature $T_0=1/2\pi L$. Thus, 
\begin{equation}
  \text{Tr}[\rho ^n]=\frac{Z(T_0/n)}{Z(T_0)^n},
\end{equation}
from which the R\'{e}nyi entropies can be calculated from the thermodynamics of the CFT living on the hyperbolic cylinder $\mathbb{R} \times \mathbb{H}^{d-1}$. In terms of the AdS/CFT correspondence, the partition function $Z(T_0/n)$ is identified with that of a hyperbolic black hole at temperature $T=T_0/n$. Thus, the R\'{e}nyi entropies can be calculated by means of the hyperbolic black hole. Furthermore, with the free energy $F=-T\log Z$ of black holes, we can rewrite the R\'{e}nyi entropies as
\begin{equation}
  S_n=\frac{n}{1-n}\frac{1}{T_0}\bigl[F(T_0)-F(T_0/n)\bigr].
  \label{eq:Sn-F}
\end{equation}
Equivalenty, with the relation $S_{\text{therm}}=-\partial F/\partial T$, we can obtain
\begin{equation}
  S_n=\frac{n}{n-1}\frac{1}{T_0}\int_{T_0/n}^{T_0}{S_{\text{therm}}(T)dT}.
  \label{eq:Sn-S}
\end{equation}

\section{Holographic R\'{e}nyi entropies from hyperbolic AdS\texorpdfstring{$_4$}{b} black holes}
\label{sec:ads4-renyi}
\subsection{Neutral hyperbolic black holes with scalar hair in AdS\texorpdfstring{$_4$}{s} spacetime}
\label{sec:ads4-neutral BH}

To study holographic R\'{e}nyi entropies, we employ a class of neutral hyperbolic black holes with scalar hair \cite{Ren:2019lgw} as a one-parameter family generalization of the MTZ black hole \cite{Martinez:2004nb}. The action is
\begin{equation}
  S=\int d^4x\sqrt{-g}\Bigl[R-\frac{1}{2}(\partial\phi)^2-V(\phi)\Bigr],
\label{eq:action4}
\end{equation}
with the potential (first found in \cite{Gao:2004tu}) 
\begin{equation}
  V(\phi)=-\frac{2}{(1+\alpha^{2})^2L^2}\Bigl[\alpha^{2}\left(3\alpha^{2}-1\right)e^{-\phi/ \alpha}+8\alpha^{2}e^{\left(\alpha-1/ \alpha\right)\phi /2}+\left(3-\alpha^{2}\right)e^{\alpha\phi}\Bigr],
\label{eq:3-exp-V}
\end{equation}
where $\alpha$ is a parameter, and the potential equals the cosmological constant when $\alpha=0$ or $\alpha \rightarrow \infty$. A remarkable feature of this potential is that it interpolates among four scalar potentials in supergravity: the values $\alpha=0,\ 1/\sqrt{3},\ 1$, and $\sqrt{3}$ correspond to special cases of STU supergravity. Since the $\phi \rightarrow 0$ behavior is $V(\phi)=-6/L^2-(1/L^2)\phi^2+\mathcal{O}(\phi^4)$, the mass of the scalar field satisfies $m^2 L^2=-2$. From the relation $m^2 L^2=\Delta(\Delta-d)$, where $d=3$, the scaling dimension of the dual operator in the CFT is $\Delta_{+}=2$ or $\Delta_{-}=1$.\footnote{The asymptotic behavior of the scalar field near the AdS boundary is determined by its mass. If we use Fefferman-Graham (FG) coordinates, the asymptotic expansion is $\phi(x,z)=z^{\Delta_{-}}(\phi_a+\phi_bz+\cdots)$ for the alternative quantization. We consider a sourceless boundary condition corresponding to a triple-trace deformation in the dual CFT \cite{Witten:2001ua}. Specifically, the boundary condition for the following solution is given by $\phi_b/\phi_a^2=\tau$, where $\tau=-(1-\alpha^2)/4\alpha$.}

The solution to the system is \cite{Ren:2019lgw}
\begin{equation}
  ds^2=-f(r)dt^2+\frac{dr^2}{f(r)}+U(r)d \Sigma_2^2\ ,\qquad e^{\alpha\phi}=\biggl(1-\frac{b}{r}\biggr)^{\frac{2\alpha^2}{1+\alpha^2}},
\end{equation}
with
\begin{equation}
  f(r)=-\biggl(1-\frac{b}{r}\biggr)^{\frac{1-\alpha^2}{1+\alpha^2}}+\frac{r^2}{L^2}\left(1-\frac{b}{r}\right)^{\frac{2\alpha^2}{1+\alpha^2}}, \qquad U(r)=r^2\left(1-\frac{b}{r}\right)^{\frac{2\alpha^2}{1+\alpha^2}}.
\label{eq:sol-fU}
\end{equation}
This solution was obtained by taking a nontrivial neutral limit\footnote{There is also a trivial neutral limit, which gives the SAdS black hole \cite{Ren:2019lgw}.} from a class of charged dilaton black hole solutions in \cite{Gao:2004tu}. The horizon radius of black holes is obtained by taking $f(r_h)=0$, and the parameter $b$ can be replaced by $r_h$ with
\begin{equation}
  b=r_h\biggl[1-\Bigl(\frac{r_h}{L}\Bigr)^{\frac{2(1+\alpha^2)}{1-3\alpha^2}}\biggr].
\end{equation}

\begin{figure}
  \centering
  \includegraphics[width=0.49\linewidth]{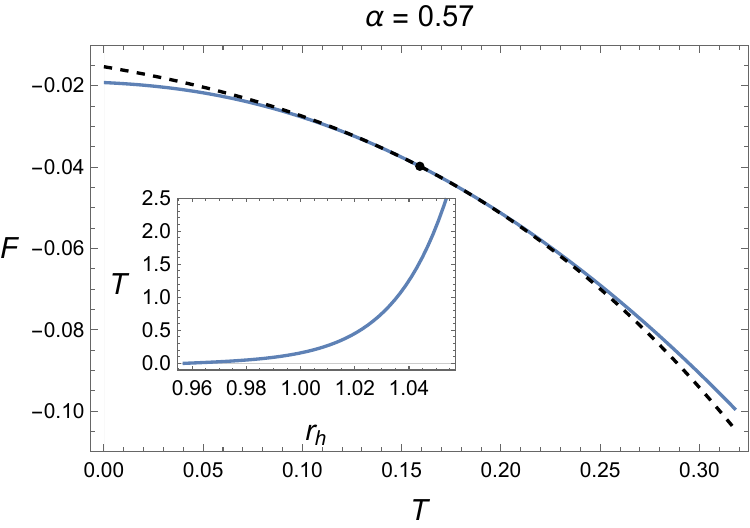}
  \includegraphics[width=0.49\linewidth]{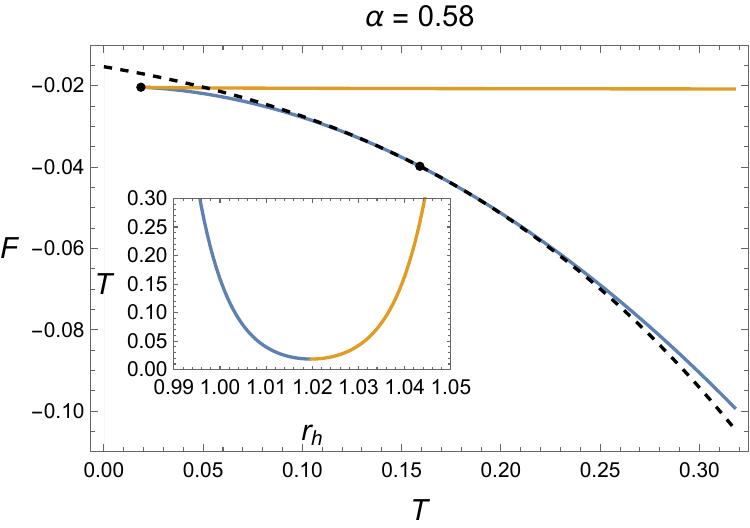}
  \includegraphics[width=0.49\linewidth]{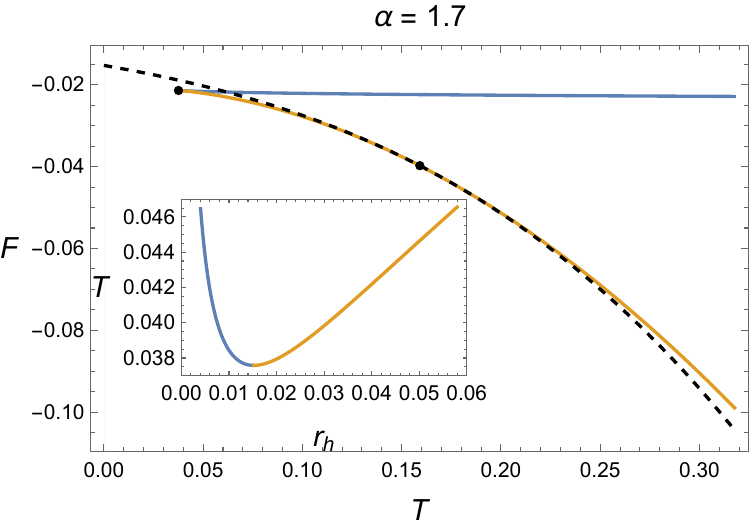}
  \includegraphics[width=0.49\linewidth]{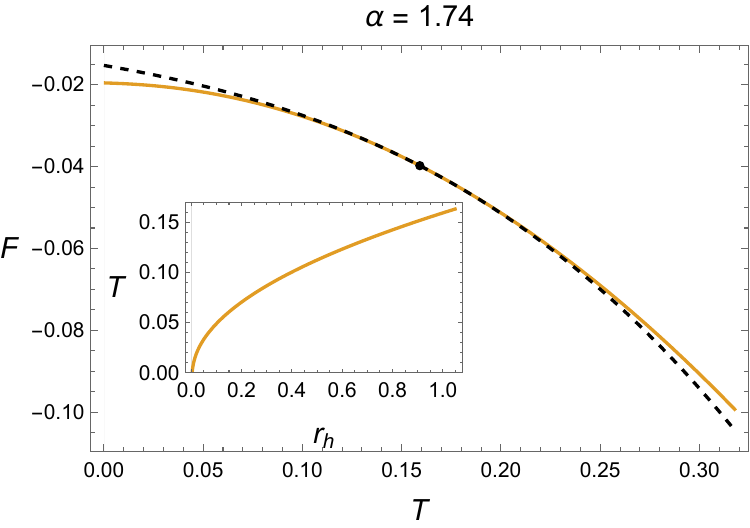}
  \caption{The free energy as a function of temperature for different values of $\alpha$. The dashed line is for the SAdS black hole, while the solid line is for the hairy black hole, where the blue one is for $r_{h,+}$ and the orange one is for $r_{h,-}$. The subfigure describes the relation between the temperature and horizon radius. In all figures, we have set $L=G=V_\Sigma=1$.}
  \label{fig:ads4-free}
\end{figure}

Now we refine the analysis of the black hole thermodynamics studied in \cite{Ren:2019lgw}. The temperature is
\begin{equation}
  T=\frac{f^{\prime}(r_h)}{4\pi}=\frac{1}{4\pi (1+\alpha^2)L}\biggl[(3-\alpha^2)\Bigl(\frac{r_h}{L}\Bigr)^\frac{1+\alpha^2}{1-3\alpha^2}-(1-3\alpha^2)\Bigl(\frac{r_h}{L}\Bigr)^{-\frac{1+\alpha^2}{1-3\alpha^2}}\biggr],
  \label{eq:T}
\end{equation}
from which we can solve $r_h$ in terms of temperature as 
\begin{equation}
  r_{h, \pm}=\Biggl(\frac{2 \pi LT(1+\alpha ^2) \pm \sqrt{4\pi^2 L^2T^2(1+\alpha^2)^2+(3-\alpha ^2) (1-3 \alpha ^2)}}{3-\alpha ^2}\Biggr)^{\frac{1-3\alpha^2}{1+\alpha ^2}}L,
  \label{eq:ads4-rh-T}
\end{equation}
which shows that there may be two solutions of black holes at a given temperature, and the horizon radius is distinguished by $r_{h,\pm}$. More specifically, when $0\leq\alpha\leq 1/\sqrt{3}$, only $r_{h,+}$ is real; when $\alpha\geq \sqrt{3}$, only $r_{h,-}$ is real; when $1/\sqrt{3}<\alpha<\sqrt{3}$, both $r_{h,\pm}$ are real. The two solutions can be seen in figure~\ref{fig:ads4-free}. The temperature can reach zero when $\alpha \leq 1/\sqrt{3}$ or $\alpha \geq \sqrt{3}$. However, when $1/\sqrt{3}<\alpha<\sqrt{3}$, there is a minimum temperature above zero,
\begin{equation}
  T_{\text{m}}=\frac{\sqrt{(3-\alpha ^2) (3 \alpha ^2-1)}}{2\pi (1+\alpha ^2) L},
  \label{eq:Tc-ads4}
\end{equation}
at which the two solutions intersect. In a special case $\alpha=1$, the $T_{\text{m}}$ has a maximum value $T_0=1/2\pi L$.

The Bekenstein-Hawking entropy is given by
\begin{equation}
  S=\frac{V_\Sigma}{4G}U(r_h)=\frac{V_\Sigma}{4G} \Bigl(\frac{r_{h}}{L}\Bigr)^{\frac{2(1-\alpha^2)}{1-3\alpha^2}}L^2,
  \label{eq:S-ads4}
\end{equation}
where $V_\Sigma$ is the (regulated) area of the hyperbolic space $d \Sigma_{2}^2$. With a boundary condition corresponding to a triple-trace deformation in the dual CFT \cite{Ren:2019lgw,Caldarelli:2016nni}, the mass derived from the holographic renormalization is
\begin{equation}
  M=-\frac{V_\Sigma}{8\pi G}\frac{1-\alpha^2}{1+\alpha^2}b=-\frac{V_\Sigma}{8\pi G}\frac{1-\alpha^2}{1+\alpha^2} r_h\biggl[1-\Bigl(\frac{r_h}{L}\Bigr)^{\frac{2(1+\alpha^2)}{1-3\alpha^2}}\biggr].
\end{equation}
All thermodynamic quantities above can be expressed as functions of temperature by means of \eqref{eq:ads4-rh-T}. So we can easily verify the first law of thermodynamics in the canonical ensemble: $dM=TdS$. Finally, the Helmholtz free energy is 
\begin{equation}
  F=M-TS=-\frac{V_{\Sigma}}{16\pi G}r_{h} \biggl[1+\Bigl(\frac{r_{h}}{L}\Bigr)^{\frac{2(1+\alpha ^2)}{1-3 \alpha ^2}}\biggr].
  \label{eq:F-ads4}
\end{equation}
A more rigorous treatment of the black hole thermodynamics is in appendix~\ref{sec:bh-thermo}, in which the boundary counter terms are given together with the boundary condition of the scalar field.

The stability of black holes can be analyzed by comparing the free energy of black holes with and without scalar hair. Note that when $\alpha=0$, the black hole is the SAdS black hole. For representative cases of $\alpha$, we plot the free energy as a function of temperature in figure~\ref{fig:ads4-free}. The SAdS black hole is always more stable than the hairy black hole at $T > T_0$. As the temperature is decreased, we draw the following conclusions on phase transitions:
\begin{itemize}
  \item $0 < \alpha <1$: The SAdS black hole experiences a third-order phase transition to the hairy one with the horizon size $r_{h,+}$ at $T=T_0$.
  \begin{itemize}
    \item[$\vcenter{\hbox{\scalebox{0.8}{\textbullet}}}$] $0 < \alpha \leq 1/\sqrt{3}$: There is only one hairy black hole corresponding to $r_{h,+}$.
    \item[$\vcenter{\hbox{\scalebox{0.8}{\textbullet}}}$] $1/\sqrt{3} < \alpha < 1$: The stable hairy black hole corresponds to $r_{h,+}$, while an unstable branch corresponds to $r_{h,-}$. A zeroth-order phase transition to the SAdS black hole occurs at $T_{\text{m}}$.
  \end{itemize}
  \item $\alpha=1$: The SAdS black hole is always more stable than the hairy black hole.
  \item $\alpha>1$: The SAdS black hole experiences a third-order phase transition to the hairy one with the horizon size $r_{h,-}$ at $T=T_0$.
  \begin{itemize}
    \item[$\vcenter{\hbox{\scalebox{0.8}{\textbullet}}}$] $1<\alpha<\sqrt{3}$: The stable hairy black hole corresponds to $r_{h,-}$, while an unstable branch corresponds to $r_{h,+}$. A zeroth-order phase transition to the SAdS black hole occurs at $T_{\text{m}}$.
    \item[$\vcenter{\hbox{\scalebox{0.8}{\textbullet}}}$] $\alpha \geq \sqrt{3}$: There is only one hairy black hole corresponding to $r_{h,-}$
  \end{itemize}
\end{itemize}

\subsection{Holographic R\'{e}nyi entropies}
We calculate the R\'{e}nyi entropies from the free energy by~\eqref{eq:Sn-F} (the thermal entropy calculation~\eqref{eq:Sn-S} gives the same result). The expression~\eqref{eq:Sn-F} also describes the R\'{e}nyi entropies with phase transitions but should be treated carefully. We need to take the thermodynamically preferred solution to calculate the free energy and entropy at temperature $T_0$ and $T_0/n$. Notably, the R\'{e}nyi entropies with $n<1$ for any $\alpha$ are the same as those from the SAdS black hole, since the most stable solution is always the SAdS black hole when $T>T_0$ (see figure~\ref{fig:ads4-free}).

We first obtain the R\'{e}nyi entropies calculated from the hairy black holes without taking into account phase transitions to the SAdS black hole,
\begin{equation}
  S_n=\frac{n L^2}{8(n-1)G}\Bigl[2-\mathcal{S}_{\pm}(\alpha,n)^{\frac{1-3\alpha^2}{1+\alpha^2}}-\mathcal{S}_{\pm}(\alpha,n)^{\frac{3-\alpha^2}{1+\alpha^2}}\Bigr]V_{\Sigma}\,,
  \label{eq:ads4-renyi-general}
\end{equation}
where
\begin{equation}	
  \mathcal{S}_{\pm}(\alpha,n)=\frac{1}{(3-\alpha ^2) n}\Bigl[(1+\alpha ^2)\pm \sqrt{(1+\alpha ^2)^2+(3-\alpha ^2) (1-3 \alpha ^2) n^2}\Bigr].
  \label{eq:ads4-renyi-co}	
\end{equation}
Here $\mathcal{S}_{\pm}$ correspond to the two solutions $r_{h,\pm}$: $\mathcal{S}_{+}$ is for $0\leq \alpha \leq 1$ and $\mathcal{S}_{-}$ is for $\alpha \geq 1$. Clearly, $S_n$ is an analytic function of $n$ with a parameter $\alpha$. The R\'{e}nyi entropies from the SAdS black hole are given by $S_n(\alpha=0)$.

We use $\bar{S}_n$ to denote the R\'{e}nyi entropies with phase transitions, and it is non-analytic at $n$ corresponding to the phase transition temperature of the black holes, $T=T_0/n=(2\pi Ln)^{-1}$. When $0<\alpha \leq 1/\sqrt{3}$ and $\alpha \geq \sqrt{3}$, the R\'{e}nyi entropies are given by
	\begin{equation}	
  \bar{S}_n=	
  \begin{cases}	
    S_n(\alpha=0),\quad &n\leq 1,\\	
    S_n,&n>1,
  \end{cases}	
  \label{eq:Sn-phase-tran1}	
\end{equation}
where $S_n$ are the R\'{e}nyi entropies from hairy black holes given by~\eqref{eq:ads4-renyi-general}. The phase transition occurs at $n=1$ ($T=T_0$), at which $\bar{S}_n$ is continuous, while its second derivative $\partial_n^2\bar{S}_n$ is discontinuous. When $1/\sqrt{3}<\alpha<\sqrt{3}$, the R\'{e}nyi entropies are given by
	\begin{equation}	
  \bar{S}_n=	
  \begin{cases}	
    S_n(\alpha=0),\quad &n\leq 1\, \text{ or }\, n\geq n_\text{m},\\	
    S_n,&1<n<n_\text{m},	
  \end{cases}	
  \label{eq:Sn-phase-tran2}	
\end{equation}
where $n_\text{m}$ is the critical parameter of the R\'{e}nyi entropies derived from~\eqref{eq:Tc-ads4},
\begin{equation}
  n_{\text{m}}=\frac{T_0}{T_{\text{m}}}=\frac{1+\alpha ^2}{\sqrt{(3-\alpha ^2) (3 \alpha ^2-1)}} \geq 1.
\end{equation}
Besides the phase transition at $n=1$, the R\'{e}nyi entropies $\bar{S}_n$ are discontinuous at $n=n_\text{m}$.

From \eqref{eq:ads4-renyi-co}, we have $\lim_{n \to 1}{\mathcal{S}_{\pm}(\alpha,n)}=1$, so the $n\rightarrow 1$ limit of the R\'{e}nyi entropies (for any $\alpha$) gives 
\begin{equation}
  S_{EE}=\frac{L^2}{4G} V_{\Sigma}\,,
\end{equation}
which means that the scalar field does not affect the entanglement entropy ($n=1$), while it affects the R\'{e}nyi entropies with $n>1$. Moreover, the R\'{e}nyi entropies also satisfy an area law \cite{Dong:2016fnf}
\begin{equation}
  n^2 \partial_{n}\Bigl(\frac{n-1}{n}S_n \Bigr)=\frac{\text{Area}(\text{Brane}_n)}{4G}=\frac{L^2}{4G}\mathcal{S}_{\pm}(\alpha,n)^{\frac{2 (1-\alpha ^2)}{1+\alpha^2}} V_{\Sigma}\,,
  \label{eq:cosmic-brane}
\end{equation}
where the $\text{Brane}_n$ is a bulk codimension-2 cosmic brane homologous to the entangling region A,\footnote{Here A is the region enclosed by the spherical entangling surface.} and the brane tension is $T_n=(n-1)/4nG$. Since $\lim_{n \to 1}{\mathcal{S}_{\pm}(\alpha,n)}=1$ and $\lim_{n\to 1}{T_n}=0$, the one-parameter generalized area law will reduce to the RT formula in the $n\to 1$ limit. It is straightforward to verify that \eqref{eq:cosmic-brane} is exactly the entropy of the hyperbolic black hole.

\begin{figure}	
  \centering	
  \includegraphics[width=0.49\linewidth]{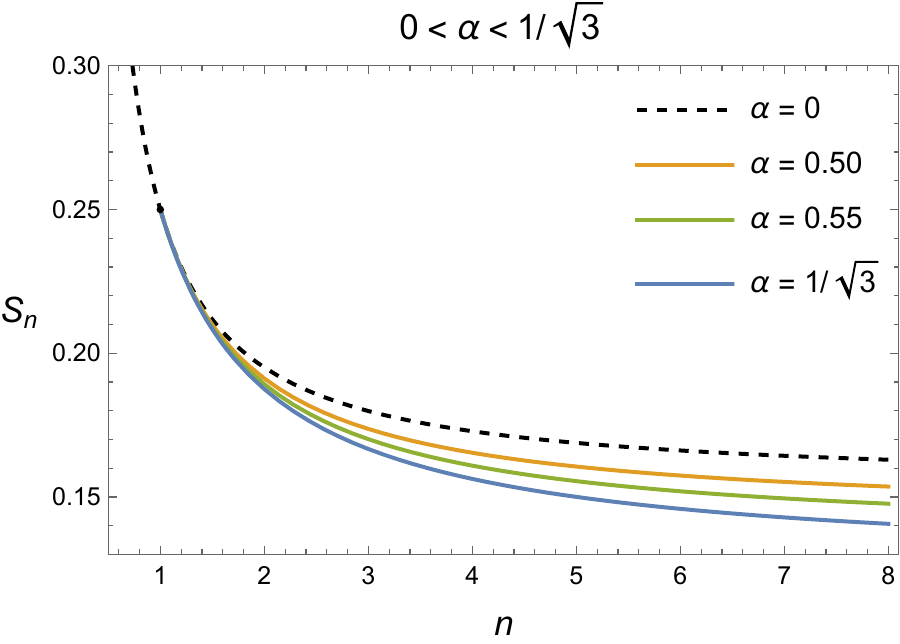}	
  \includegraphics[width=0.49\linewidth]{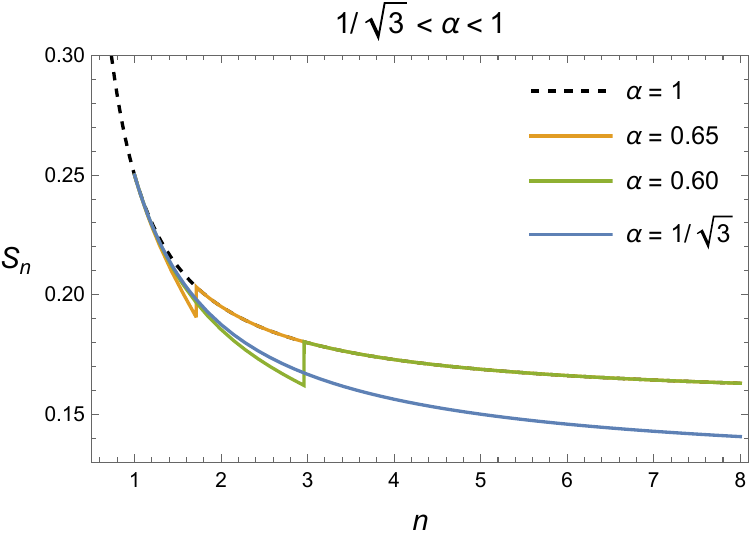}	
  \includegraphics[width=0.49\linewidth]{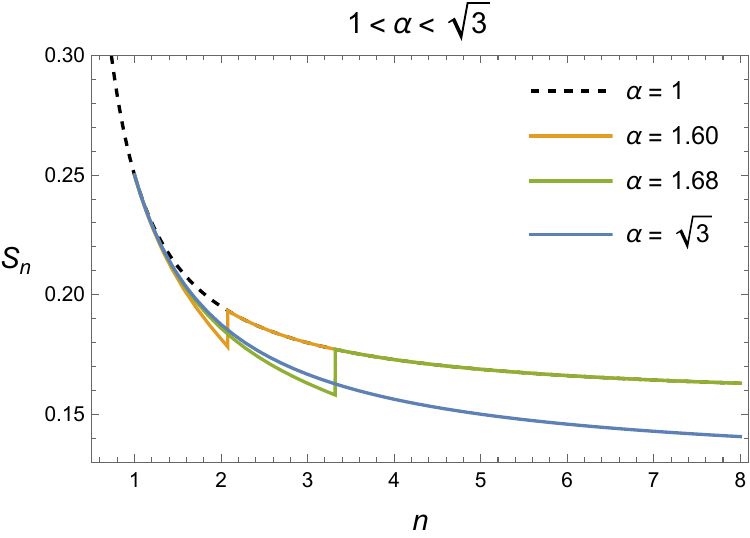}	
  \includegraphics[width=0.49\linewidth]{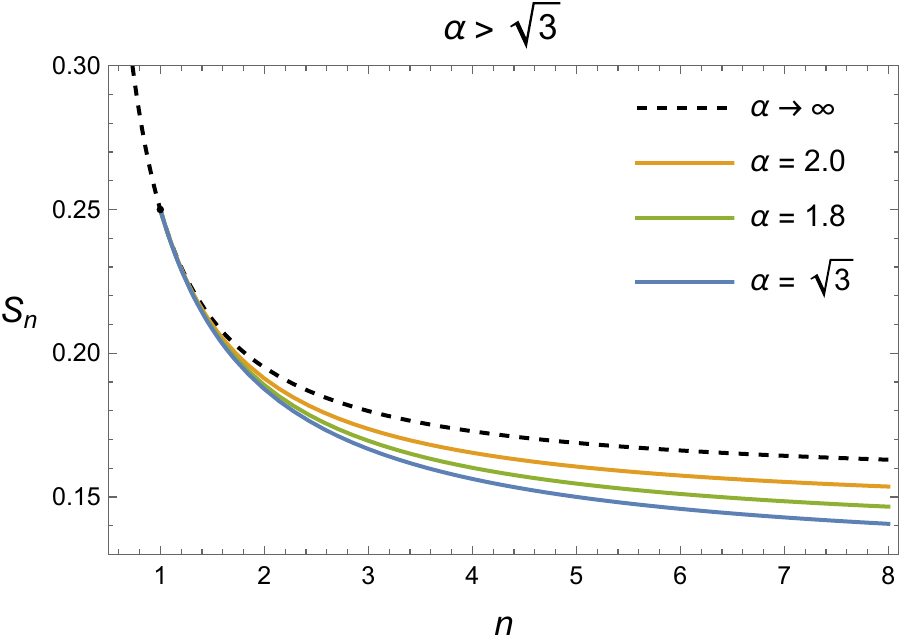}	
  \caption{The R\'{e}nyi entropies as a function of $n$ for different values of $\alpha$ (AdS$_4$). The dashed line is for the SAdS$_4$ case, and the blue line is for $\alpha=1/\sqrt{3}$ or $\sqrt{3}$. A third-order phase transition occurs at $n=1$ (for all $\alpha \neq 1$), where the second derivative has discontinuity. When $1/\sqrt{3}<\alpha<\sqrt{3}$, a zeroth-order phase transition of black holes leads to discontinuity of the R\'{e}nyi entropies at $n_\text{m}>1$.}	
  \label{fig:ads4-renyi}	
\end{figure}

The R\'{e}nyi entropies as a function of $n$ for different values $\alpha$ are plotted in figure~\ref{fig:ads4-renyi}. The most notable cases are when $\alpha$ takes values corresponding to special cases of STU supergravity.\footnote{There are U(1)$^4$ gauged fields in STU supergravity for AdS$_4$ spacetime \cite{Cvetic:1999xp}, with four U(1) charges, respectively. Here the special values of $\alpha$ correspond to the cases when some of the charges are the same while others are zero (see \cite{Ren:2019lgw} for more details).} We use the same terminology as in \cite{Ren:2019lgw}: 
\begin{itemize}
  \item 1-charge black hole ($\alpha=\sqrt{3}$) \\
  This black hole is exactly the MTZ black hole. The free energy is
  \begin{equation}
    F=-\frac{L}{16\pi G}(1+4\pi^2 L^2 T^2)V_{\Sigma}\,.
    \label{eq:F-sqrt3}
  \end{equation}
  From~\eqref{eq:Sn-F}, the R\'{e}nyi entropies are 
  \begin{equation}
    S_n=\frac{L^2}{8G}\biggl(1+\frac{1}{n}\biggr) V_{\Sigma}\,,
    \label{eq:Sn-sqrt3}
  \end{equation}
  for $n \geq 1$. Note that the result has the same $n$ dependence as the universal result in 2D CFTs with a single interval \cite{Calabrese:2004eu},
  \begin{equation}
    S_n(D=2)=\frac{c}{6}\biggl(1+\frac{1}{n}\biggr)\log \frac{l}{\delta}\,,
    \label{eq:Sn-2dcft}
  \end{equation}
  where $c$ is the central charge, $\delta$ is a short-distance cut-off, and $l$ is the size of the interval. Besides, \eqref{eq:Sn-sqrt3} or \eqref{eq:Sn-2dcft} can also be derived from the SAdS$_3$ hyperbolic black hole \cite{Hung:2011nu}.

  \item 2-charge black hole ($\alpha=1$) \\
  In this case, the free energy is linear with the temperature 
  \begin{equation}
    F=-\frac{L^2}{4G} T V_{\Sigma}\,,
  \end{equation}
  which is always less stable than the SAdS black hole. So the R\'{e}nyi entropies are $S_n=S_n(\alpha=0)$.

  \item 3-charge black hole ($\alpha=1/\sqrt{3}$) \\
The system \eqref{eq:action4}--\eqref{eq:sol-fU} has an invariance under $\alpha \rightarrow 1/\alpha$ and $b\to -b$. So the free energy and R\'{e}nyi entropies are the same as \eqref{eq:F-sqrt3} and \eqref{eq:Sn-sqrt3}, respectively. As a remark, for an EMD system with the potential~\eqref{eq:3-exp-V}, the $\alpha=1/\sqrt{3}$ case is also called the Gubser-Rocha model, and the extremal near-horizon geometry of its 11-dimensional lift has an AdS$_3$ factor \cite{Gubser:2009qt}.

  \item 4-charge black hole ($\alpha=0$) \\
  The black hole is the SAdS black hole. The free energy is 
  \begin{equation}
    F=-\frac{L}{108\pi G}\bigl[\pi LT(9+8 \pi ^2 L^2 T^2)+(3+4 \pi ^2 L^2 T^2)^{3/2}\bigr]V_{\Sigma}\,.
  \end{equation}
  The R\'{e}nyi entropies are
  \begin{equation}
    S_n(\alpha=0)=\frac{n L^2}{8 (n-1)G}\biggl[2-\frac{1+\sqrt{1+3 n^2}}{3 n}-\biggl(\frac{1+\sqrt{1+3n^2}}{3n}\biggr)^3\biggr]V_{\Sigma}\,,
  \end{equation}
  which exactly gives the result from the SAdS$_4$ black hole in \cite{Hung:2011nu}.
\end{itemize}

\subsection{Some R\'{e}nyi entropy inequalities}
\label{sec:inequal}
The R\'{e}nyi entropies should satisfy the following inequalities \cite{Beck:1993}
\begin{align}
  \frac{\partial S_n}{\partial n} \leq 0\label{eq:inequal1},\\
  \frac{\partial}{\partial n}\Bigl(\frac{n-1}{n}S_n\Bigr)\geq 0\label{eq:inequal2},\\
  \frac{\partial}{\partial n}\bigl((n-1)S_n\bigr)\geq 0\label{eq:inequal3},\\
  \frac{\partial^2}{\partial n^2}\bigl((n-1)S_n\bigr)\leq 0\label{eq:inequal4}.
\end{align}
From the holographic calculation, these inequalities are closely related to the stability of hyperbolic black holes (or the CFT living on the hyperbolic cylinder), which has been discussed\footnote{These inequalities can also be proved under the assumption of bulk stability against perturbation \cite{Nakaguchi:2016zqi} through the holographic description of the R\'{e}nyi entropies in \cite{Dong:2016fnf}.} in detail in \cite{Hung:2011nu} by combining these inequalities with~\eqref{eq:Sn-S}. Among them,~\eqref{eq:inequal1} and~\eqref{eq:inequal4} are ensured by the positive specific heat of black holes, while~\eqref{eq:inequal2} is guaranteed by the positivity of thermal entropy. Besides,~\eqref{eq:inequal3} is related to both specific heat and thermal entropy.  

We can verify that all these inequalities are satisfied for the R\'{e}nyi entropies that we obtained from the thermodynamically preferred branch of black hole solutions. The thermal entropy is positive from~\eqref{eq:S-ads4} since $r_{h}>0$. The specific heat can be derived from~\eqref{eq:ads4-rh-T} and~\eqref{eq:S-ads4} as
\begin{equation}	
  c=T\frac{\partial S}{\partial T}=\pm \frac{V_{\Sigma}}{G} \frac{\pi (1-\alpha^2)L^3 T}{\sqrt{4\pi^2L^2 T^2(1+\alpha^2)^2+(3-\alpha^2)(1-3\alpha^2)}} \Bigl(\frac{r_{h,\pm}}{L}\Bigr)^{\frac{2(1-\alpha^2)}{1-3\alpha^2}}.	
  \label{eq:specific-heat}	
\end{equation}
As discussed in section~\ref{sec:ads4-neutral BH}, the thermodynamically preferred solution is either the SAdS black hole or one of the hairy branches, which depends on the values of $\alpha$ and the R\'{e}nyi parameter $n$. The specific heat of the SAdS solution is obvious to see by taking $\alpha=0$ ($r_{h,+}$). Between the two hairy black hole branches, the one with $r_{h,+}$ is more stable when $0<\alpha<1$, while the one $r_{h,-}$ is more stable when $\alpha>1$. Thus the specific heat is always positive for the more stable hairy solution. When $\alpha=1$, the hairy black hole is unstable with zero specific heat, so the black hole is always SAdS. Therefore, these R\'{e}nyi entropy inequalities hold for all values of $\alpha$ (except for the critical points where the derivatives of the R\'{e}nyi entropies are discontinuous).

\section{Holographic R\'{e}nyi entropies from hyperbolic AdS$_5$ black holes}
\label{sec:ads5-renyi}
It is natural to generalize the AdS$_4$ cases to higher dimensions, especially AdS$_5$, whose special cases also correspond to supergravity. A class of black hole solutions to the EMD system in higher dimensions were found in \cite{Gao:2004tv,Gao:2005xv}. We still focus on the nontrivial neutral limit \cite{Ren:2019lgw}.
\subsection{Neutral hyperbolic black holes with scalar hair in AdS\texorpdfstring{$_5$}{s} spacetime}
The action in AdS$_5$ is 
\begin{equation}
  S=\int d^5x\sqrt{-g}\Bigl[R-\frac{1}{2}(\partial\phi)^2-V(\phi)\Bigr],
\end{equation}
with the potential\footnote{The expansion of the potential near $\phi \rightarrow 0$ is $V(\phi)=-12/L^2-2\phi^2/L^2+\mathcal{O}(\phi^3)$, so the mass of the scalar field is $m^2L^2=-4$ and the scaling dimension of the dual operator is $\Delta_{\pm}=2$. The scalar field behaves as $\phi(r) \sim \phi_a\ln{r}/r^2+\phi_b/r^2$ near the AdS boundary. The boundary condition for the following solution corresponds to the standard quantization $\phi_a=0$.}
\begin{equation}
  V(\phi)=-\frac{12}{(4+3 \alpha ^2)^2 L^2} \Bigl[3\alpha ^2 (3 \alpha ^2-2) e^{-\frac{4 \phi }{3 \alpha }}+36 \alpha ^2 e^{\frac{(3 \alpha ^2-4) \phi }{6 \alpha }}+2 (8-3 \alpha ^2) e^{\alpha  \phi }\Bigr].
\end{equation}
The solution is \cite{Ren:2019lgw}
\begin{equation}
  ds^2=-f(r)dt^2+\frac{dr^2}{g(r)}+U(r)d\Sigma_3^2\ ,\qquad e^{\alpha \phi }=\biggl(1-\frac{b^2}{r^2}\biggr)^{\frac{6 \alpha ^2}{4+3 \alpha ^2}},
\end{equation}
with
\begin{align}
  \begin{split}
    f(r)&=-\biggl(1-\frac{b^2}{r^2}\biggr)^{\frac{4-3 \alpha ^2}{4+3 \alpha ^2}}+\frac{r^2 }{L^2}\biggl(1-\frac{b^2}{r^2}\biggr)^{\frac{3 \alpha ^2}{4+3 \alpha ^2}},\\
    g(r)&=f(r) \biggl(1-\frac{b^2}{r^2}\biggr)^{\frac{3 \alpha ^2}{4+3 \alpha ^2}},\qquad U(r)=r^2 \biggl(1-\frac{b^2}{r^2}\biggr)^{\frac{3 \alpha ^2}{4+3 \alpha ^2}}.
  \end{split}
\end{align}

From $f(r_h)=0$, the parameter $b^2$ can be replaced with 
\begin{equation}
  b^2=r_h^2\Bigl[1-\biggl(\frac{r_h}{L}\Bigr)^{\frac{4+3 \alpha ^2}{2-3 \alpha ^2}}\biggr],
\end{equation}
and thus the temperature is
\begin{equation}
  T=\frac{\sqrt{f'(r_h)g'(r_h)}}{4\pi}=\frac{1}{2 \pi (4+3 \alpha ^2) L}\biggl[(8-3 \alpha ^2)\Bigl(\frac{r_h}{L}\Bigr)^{\frac{4+3\alpha ^2}{4-6 \alpha ^2}}-(4-6 \alpha ^2)\Bigl(\frac{r_h}{L}\Bigr)^{-\frac{4+3 \alpha^2}{4-6\alpha ^2}}\biggr].
\end{equation}
So we have
\begin{equation}
  r_{h, \pm}=\Biggl(\frac{\pi LT(4+3 \alpha ^2) \pm \sqrt{\pi ^2 L^2T^2 (4+3 \alpha ^2)^2+(8-3 \alpha ^2) (4-6 \alpha ^2)}}{8-3 \alpha ^2}\Biggr)^{\frac{4-6 \alpha ^2}{4+3 \alpha ^2}}L.
  \label{eq:ads5-rh-T}
\end{equation}
Other thermodynamic quantities are
\begin{align}
  M&=-\frac{3 V_{\Sigma }}{16 \pi  G} \frac{4-3 \alpha ^2}{4+3 \alpha ^2} r_h^2\biggl[1-\Bigl(\frac{r_h}{L}\Bigr)^{\frac{4+3 \alpha ^2}{2-3 \alpha ^2}}\biggr],\\
  S&=\frac{V_{\Sigma}}{4G}U(r_h)^{3/2}=\frac{V_{\Sigma } }{4 G} \Bigl(\frac{r_h}{L}\Bigr)^{\frac{3 (4-3 \alpha ^2)}{2 (2-3 \alpha ^2)}}L^3,\\
  F&=M-TS=-\frac{V_{\Sigma } }{16 \pi  G}r_h^2 \biggl[1+ \Bigl(\frac{r_h}{L}\Bigr)^{\frac{4+3 \alpha ^2}{2-3 \alpha ^2}}\biggr],
\end{align}
from which we can verify the phase transitions and the R\'{e}nyi entropy inequalities as in section~\ref{sec:ads4-renyi}.

Similar to the discussion in AdS$_4$ cases, \eqref{eq:ads5-rh-T} shows that there may be two solutions of black holes at a given temperature: $r_{h,+}$ is real when $0\leq \alpha < 4/\sqrt{6}$, while $r_{h,-}$ is real when $\alpha > 2/\sqrt{6}$. The temperature can reach zero when $0 \leq \alpha \leq 2/\sqrt{6}$ or $\alpha \geq 4/\sqrt{6}$, while there is a minimum temperature when $2/\sqrt{6}<\alpha<4/\sqrt{6}$,
\begin{equation}
  T_{\text{m}}=\frac{\sqrt{2 (3 \alpha ^2-2) (8-3 \alpha ^2)}}{\pi (3 \alpha ^2+4) L},
  \label{eq:ads5-Tc}
\end{equation}  
at which a zeroth-order phase transition from hairy black holes to the SAdS black hole occurs. The maximum value of $T_\text{m}$ is $T_0=1/2\pi L$ in the case $\alpha=2/\sqrt{3}$. Besides, there is a third-order phase transition from the SAdS hole to hairy black holes at $T=T_0$ for any $\alpha \neq 2/\sqrt{3}$.  

\subsection{Holographic R\'{e}nyi entropies}
The calculation of the R\'{e}nyi entropies is similar to that in section~\ref{sec:ads4-renyi}. We first obtain the result from hairy black holes without taking in to account phase transitions,
\begin{equation}
  S_n=\frac{n L^3}{8(n-1)G}\Bigl[2-\mathcal{S}_{\pm}(\alpha,n)^{\frac{8-12\alpha^2}{4+3\alpha^2}}-\mathcal{S}_{\pm}(\alpha,n)^{\frac{16-6\alpha^2}{4+3\alpha^2}}\Bigr] V_{\Sigma}\,,
  \label{eq:ads5-Sn-general}
\end{equation}
where
\begin{equation}	
  \mathcal{S}_{\pm}(\alpha,n)=\frac{1}{2(8-3\alpha^2)n}\Bigl[(4+3\alpha^2)\pm \sqrt{(4+3\alpha^2)^2+8(8-3\alpha^2) (2-3\alpha^2)n^2}\Bigr],	
\end{equation}
where the $\pm$ signs correspond to $r_{h,\pm}$ similar to AdS$_4$ cases. With phase transitions, the R\'{e}nyi entropies $\bar{S}_n$ is non-analytic at $n$ corresponding to the phase transition temperature of the black holes. When $0<\alpha \leq 2/\sqrt{6}$ and $\alpha \geq 4/\sqrt{6}$, the R\'{e}nyi entropies $\bar{S}_n$ are given by~\eqref{eq:Sn-phase-tran1}, in which $S_n$ is \eqref{eq:ads5-Sn-general}. When $2/\sqrt{6}<\alpha<4/\sqrt{6}$, the R\'{e}nyi entropies $\bar{S}_n$ are given by~\eqref{eq:Sn-phase-tran2} with the critical parameter
\begin{equation}
  n_{\text{m}}=\frac{T_0}{T_{\text{m}}}=\frac{4+3\alpha^2}{\sqrt{2 (3 \alpha ^2-2) (8-3 \alpha ^2)}}.
\end{equation}

\begin{figure}
  \centering
  \includegraphics[width=0.49\linewidth]{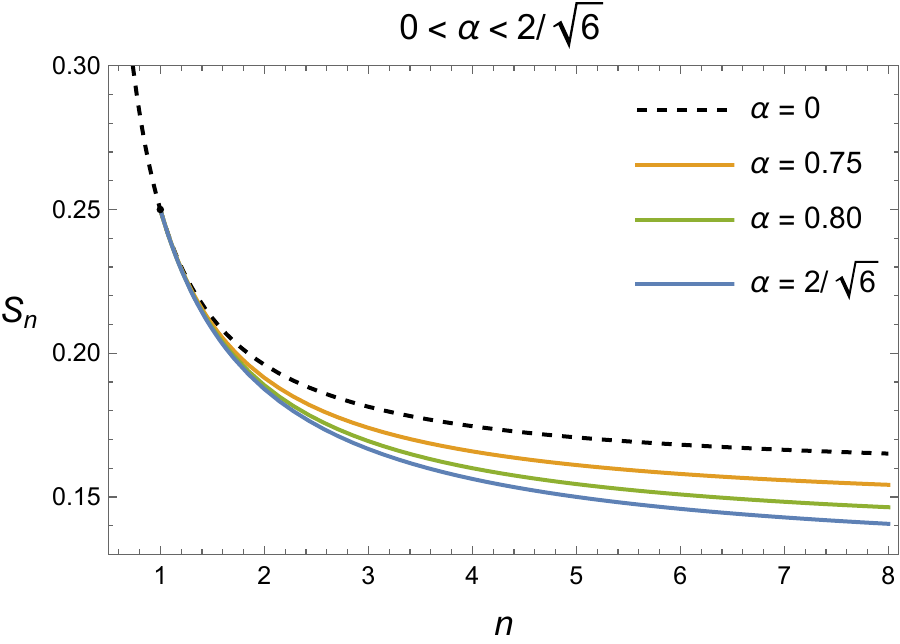}
  \includegraphics[width=0.49\linewidth]{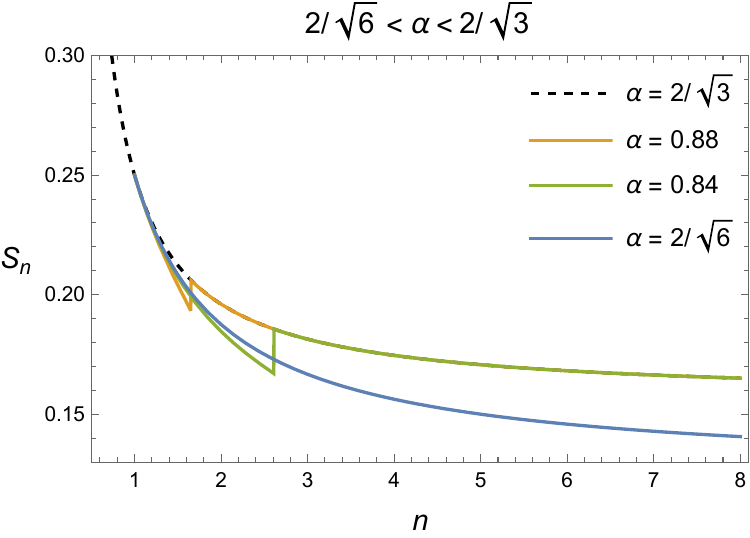}
  \includegraphics[width=0.49\linewidth]{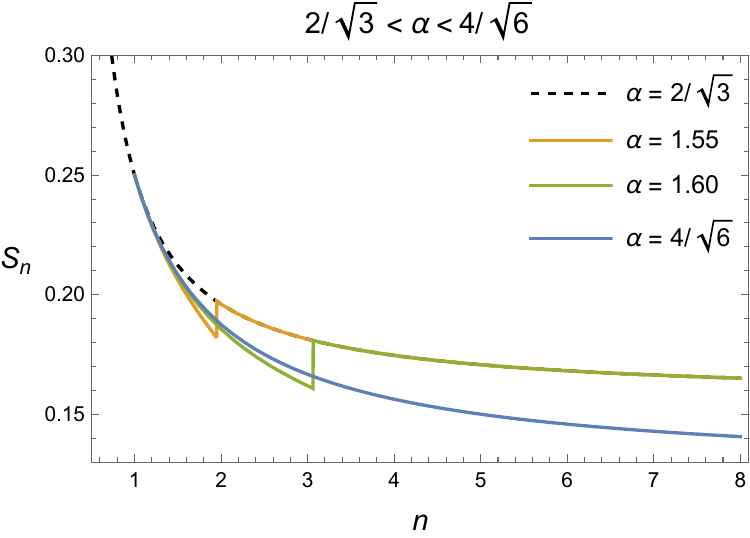}
  \includegraphics[width=0.49\linewidth]{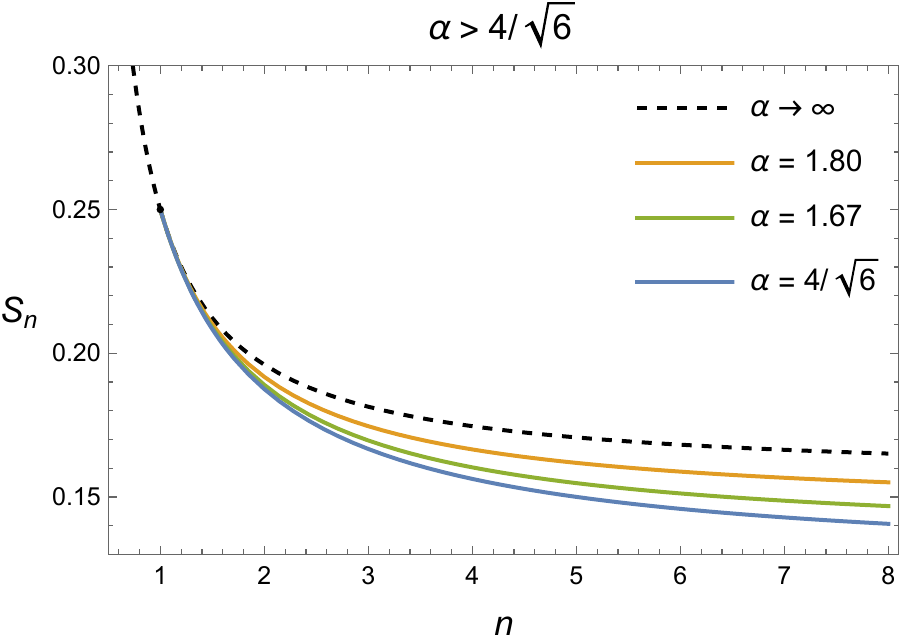}
  \caption{The R\'{e}nyi entropies as a function of $n$ for different values of $\alpha$ (AdS$_5$). The dashed line is for the SAdS$_5$ case, and the blue line is for $\alpha=2/\sqrt{6}$ or $4/\sqrt{6}$. A third-order phase transition occurs at $n=1$ (for all $\alpha \neq 2/\sqrt{3}$), where the second derivative has discontinuity. When $2/\sqrt{6}<\alpha<4/\sqrt{6}$, a zeroth-order phase transition of black holes leads to discontinuity of the R\'{e}nyi entropies at $n_\text{m}>1$.}
  \label{fig:ads5-renyi}
\end{figure}

The R\'{e}nyi entropies as a function of $n$ for different values $\alpha$ are plotted in figure~\ref{fig:ads5-renyi}. The black holes corresponding to special cases of supergravity are:\footnote{Unlike the AdS$_4$ cases with the four U(1) gauge fields, there are three U(1) gauge fields in AdS$_5$ in STU supergravity \cite{Cvetic:1999xp}, and thus we have three U(1) charges here.}
\begin{itemize}
  \item 1-charge black hole ($\alpha=4/\sqrt{6}$)\\
  We find the free energy when $\alpha=4/\sqrt{6}$ in AdS$_5$ is exactly the same as the one when $\alpha=\sqrt{3}$ in AdS$_4$. The results are
  \begin{align}
    F=-\frac{L^2}{16\pi G}(1+4\pi^2 L^2 T^2)V_{\Sigma}\,, \qquad S_n=\frac{L^3}{8G}\biggl(1+\frac{1}{n}\biggr)V_{\Sigma}\,.
    \label{eq:ads5-Sn-sqrt}
  \end{align}
  \item 2-charge black hole ($\alpha=2/\sqrt{6}$)\\
  This solution corresponds to the same system with the $1$-charge black hole~\eqref{eq:ads5-Sn-sqrt}.
  \item 3-charge black hole ($\alpha=0$)\\
  The scalar hair vanishes, and we obtain the SAdS$_5$ black hole with the free energy
  \begin{equation}
    F=-\frac{L^2}{128\pi G}\Bigl[6+4 \pi  L T \Bigl(\pi LT(3+\pi ^2 L^2 T^2)+(2+\pi ^2 L^2 T^2)^{3/2}\Bigr)\Bigr]V_{\Sigma}\,.
  \end{equation}
  And the R\'{e}nyi entropies are 
  \begin{equation}
    S_n(\alpha=0)=\frac{n L^3}{8 (n-1)G}\biggl[2-\biggl(\frac{1+\sqrt{1+8 n^2}}{4 n}\biggr)^2-\biggl(\frac{1+\sqrt{1+8 n^2}}{4 n}\biggr)^4\biggr] V_{\Sigma}\,.
  \end{equation}
\end{itemize}

The R\'{e}nyi entropies $S_n$ as a function of $n$ can be easily generalized to higher dimensions, based on the solution of hyperbolic black holes with scalar hair in AdS$_{d+1}$; see \cite{Ren:2019lgw} for details of this solution. The R\'{e}nyi entropies calculated from the hairy black holes are
\begin{equation}	
  S_n=\frac{n L^{d-1}}{8(n-1)G}\bigl[2-\mathcal{S}_{\pm}(\alpha,n)^{\mathsf{a}/\mathsf{c}}-\mathcal{S}_{\pm}(\alpha,n)^{\mathsf{b}/\mathsf{c}}\Bigr]V_{\Sigma},	
  \label{eq:adsd-renyi-general}	
\end{equation}	
where
\begin{gather}	
  \mathcal{S}_{\pm}(\alpha,n)=\frac{1}{\mathsf{b}n}\Bigl(\mathsf{c}\pm\sqrt{\mathsf{c}^2+\mathsf{a}\mathsf{b} n^2}\Bigr),	
  \label{eq:adsd-renyi-co}\\
    \begin{cases}	
    \mathsf{a}=2 (d-2)^2-d(d-1) \alpha ^2,\\	
    \mathsf{b}=(d-2) [2 d-(d-1)\alpha ^2],\\	
    \mathsf{c}=2 (d-2)+(d-1)\alpha ^2.
  \end{cases}
\end{gather}	
With phase transitions, the R\'{e}nyi entropies $\bar{S}_n$ is non-analytic at $n$ corresponding to the phase transition temperature of the black holes, $T=T_0/n=(2\pi Ln)^{-1}$. When $0<\alpha \leq (d-2)\sqrt{\frac{2}{d(d-1)}}$ and $\alpha \geq \sqrt{\frac{2d}{(d-1)}}$, the R\'{e}nyi entropies $\bar{S}_n$ are given by~\eqref{eq:Sn-phase-tran1}, in which $S_n$ is \eqref{eq:adsd-renyi-general}. When $(d-2)\sqrt{\frac{2}{d(d-1)}}<\alpha<\sqrt{\frac{2d}{(d-1)}}$, the R\'{e}nyi entropies $\bar{S}_n$ are given by~\eqref{eq:Sn-phase-tran2} with the critical parameter $n_\text{m}$ determined by the minimum temperature. A zeroth-order phase transition occurs at the minimum temperature
\begin{equation}	
  T_{\text{m}}=\frac{\sqrt{-\mathsf{a}\mathsf{b}}}{2\pi\mathsf{c}L},	
\end{equation}	
or equivalently, 	
\begin{equation}	
  n_{\text{m}}=\frac{T_0}{T_m}=\frac{\mathsf{c}}{\sqrt{-\mathsf{a}\mathsf{b}}}.	
\end{equation}
We can consider a variant of the R\'{e}nyi entropy used in \cite{Dong:2016fnf},
\begin{equation}
\widetilde{S}_n \equiv n^2 \partial_{n}\Bigl(\frac{n-1}{n}S_n \Bigr)=\frac{L^2}{4G}\mathcal{S}_{\pm}(\alpha,n)^\frac{\mathsf{a}+\mathsf{b}}{2\mathsf{c}} V_{\Sigma}\,,
\end{equation}
which satisfies an area law and equals the thermal entropy of the hyperbolic black hole.

\section{Entanglement spectrum}
\label{sec:spectrum}
The entanglement spectrum is the eigenvalue distribution of the reduced density matrix. With the R\'{e}nyi entropies of all orders, the entanglement spectrum is determined. Generally, the spectrum should include both discrete and continuous parts \cite{Hung:2011nu}. We briefly review the analysis. Assuming that the eigenvalue distribution is discrete, we can write the R\'{e}nyi entropies as 
\begin{equation}
  S_n=\frac{1}{1-n} \log \text{Tr} [\rho^n]=\frac{1}{1-n} \log \Bigl(\sum_{i}{d_i \lambda_i^n}\Bigr),
  \label{eq:Sn-discrete}
\end{equation} 
where $\lambda_i$ are the eigenvalues of the reduced density matrix satisfying $\lambda_1>\lambda_2> \cdots$, and $d_i$ is the degeneracy of $\lambda_i$. The large $n$ expansion of~\eqref{eq:Sn-discrete} gives 
\begin{equation}
  S_n= -\log \lambda_1-\frac{1}{n}\log (d_1 \lambda_1)+\mathcal{O}\Bigl[\frac{1}{n^2},\ \frac{1}{n}\Bigl(\frac{\lambda_2}{\lambda_1}\Bigr)^n\Bigr],
\end{equation}
where the last term leads to non-analyticity in $1/n$. A continuous spectrum will also lead to non-analyticity. However, the holographic calculations in sections~\ref{sec:ads4-renyi} and~\ref{sec:ads5-renyi} do not show any non-analyticity in $1/n$, and thus the spectrum should include both discrete and continuous parts. Furthermore, from the thermal perspective of the dual CFT, the analyticity allows only one discrete eigenvalue $\lambda_1$, which is also the largest eigenvalue of the continuous spectrum \cite{Hung:2011nu}. 

In terms of the analysis above, the R\'{e}nyi entropies can be written as 
\begin{equation}
  S_n=\frac{1}{1-n}\log \biggl[d_1 \lambda_1^n+\int_{0}^{\lambda_1}{\bar{\rho}(\lambda) \lambda^n d\lambda}\biggr],
  \label{eq:renyi-eigen}
\end{equation}
where $\bar{\rho}(\lambda)$ is the continuous part of the entanglement spectrum $\rho(\lambda)$. Moreover, $\bar{\rho}(\lambda)$ is analytic at $\lambda=\lambda_1$, and we can write the discrete part of $\rho(\lambda)$ as a Dirac delta function. Thus, the R\'{e}nyi entropies satisfy
\begin{equation}
  \exp [(1-n)S_n]=\int_{t_1}^{+\infty}{e^{-(n+1)t}\rho(e^{-{t}})dt},
  \label{eq:Laplace}
\end{equation}
where $\lambda$ is reparameterized as $\lambda=e^{-t}$, and $\lambda_1=e^{-t_1}$. The lower limit of the integral~\eqref{eq:Laplace} can be changed to zero, which gives nothing but the Laplace transform. Thus, the spectrum can be derived from an inverse Laplace transform with $n$ as the parameter,
\begin{equation}
  \rho(\lambda)=\frac{1}{\lambda}\mathcal{L}^{-1}\bigl[e^{(1-n)S_n},n,t \bigr] \bigl|_{t=-\log \lambda}=\frac{1}{\lambda}\frac{1}{2\pi i}\lim_{T\rightarrow \infty}\int_{\gamma-i T}^{\gamma+i T}{e^{(1-n)S_n}e^{n t}d n},
  \label{eq:inv-laplace}
\end{equation}
where the integral is taken over a vertical line with $\text{Re}(s)=\gamma$, and $\gamma$ is a real number ensuring no singularity on the right side of this line.

In condensed matter literature, people are more interested in the low-lying part\footnote{Following \cite{Haldane:2008en}, we use the term low-lying, which represents the spectrum with lower energy, or with $\lambda$ closer to the largest eigenvalue.} of the entanglement spectrum, which implies more universal properties of the system \cite{Haldane:2008en,Calabrese:2008cl}. Especially in $(1+1)$ dimensions, the strong conformal invariance fixes the continuous spectrum entirely, so the universal information beyond the entanglement entropy is embodied in the location and degeneracies of the low-lying eigenvalues \cite{Calabrese:2008cl}. However, the cases in higher dimensions imply more complex properties. The systems corresponding to different $\alpha$ have different continuous spectra as well as discrete parts.

Note that the entanglement spectrum depends on all $n$ including the $n<1$ part. It is difficult to perform the inverse Laplace transform for the R\'{e}nyi entropies $\bar{S}_n$ with phase transitions. Instead, we use the analytic solution of the R\'{e}nyi entropies $S_n$ calculated from hyperbolic black holes with scalar hair, and explicitly obtain the entanglement spectrum as an infinite sum. We show that the low-lying part of the entanglement spectrum is determined by the large $n$ behavior of the R\'{e}nyi entropies, and briefly discuss the effects of phase transitions.

\subsection{Entanglement spectrum from AdS\texorpdfstring{$_4$}{b} black holes}
\label{sec:ads4-spectrum}
We start with special cases when $\alpha=1/\sqrt{3}$ or $\sqrt{3}$, from which the spectrum can be derived straightforwardly. The R\'{e}nyi entropies have a simple form 
\begin{equation}
  S_n=\frac{L^2}{8G}\biggl(1+\frac{1}{n}\biggr)V_{\Sigma}\,.
  \label{eq:Sn-2}
\end{equation}
This result has the same $n$ dependence as the universal result~\eqref{eq:Sn-2dcft} in 2D CFTs, whose entanglement spectrum can be calculated by means of properties of the polylogarithm function \cite{Calabrese:2008cl} or Mellin's inversion formula \cite{Romero-Bermudez:2018dim}:
\begin{equation}
  \rho(\lambda)=\delta(\lambda_1-\lambda)+\frac{b\, \theta(\lambda_1-\lambda)}{\lambda\sqrt{b\ln{(\lambda_1/\lambda)}}}\, I_1\bigl(2\sqrt{b\ln{(\lambda_1/\lambda)}}\bigr),
  \label{eq:spectrum-sqrt3}
\end{equation}
with 
\begin{equation}
  b=-\ln{\lambda_1}=\frac{L^2}{8G}V_{\Sigma}\,,
\end{equation}
where $I_k(x)$ is the modified Bessel function of the first kind and $\theta(x)$ is the Heaviside step function.

In general cases, we can still calculate the entanglement spectrum in terms of an infinite sum if the R\'{e}nyi entropies are analytic at $n=\infty$. This analyticity allows us to expand the R\'{e}nyi entropies near $n=\infty$ as 
\begin{equation}
  S_{n}=\sum\limits_{i=0}^{\infty}{s_{i}n^{-i}}\,,
  \label{eq:renyi-expand-1}
\end{equation}
where the coefficients of the first two terms are\footnote{They are obtained by expanding \eqref{eq:ads4-renyi-general}, which is well defined at $n\to\infty$ when $0\leq \alpha \leq 1/\sqrt{3}$ or $\alpha \geq \sqrt{3}$. When $1/\sqrt{3}<\alpha<\sqrt{3}$, the $n\to\infty$ behavior of the R\'{e}nyi entropies is determined by the SAdS black hole ($\alpha=0$) due to the zeroth-order phase transition of the hyperbolic black holes.} 
\begin{align}
  \begin{split}
    s_0&=\frac{L^2}{4G} \Bigl(1-2\bigl\lvert1-\alpha ^2\bigr\rvert \cdot \bigl\lvert1-3\alpha^2\bigr\rvert^{\frac{1-3 \alpha ^2}{2 (1+\alpha ^2)}}\bigl\lvert3-\alpha ^2\bigr\rvert^{\frac{\alpha ^2-3}{2 (1+\alpha ^2)}}\Bigr)V_{\Sigma}\,,\\
    s_1&=s_0-\frac{L^2}{4G} \biggl(\frac{ 1-3 \alpha ^2}{3-\alpha ^2} \biggr)^{\frac{1-\alpha ^2}{1+\alpha ^2}} V_{\Sigma}\,,
  \end{split}
\end{align} 
whose limits at $\alpha= 1/\sqrt{3}$ and $\sqrt{3}$ are
\begin{equation}
  s_0=s_1=\frac{L^2}{8G}V_{\Sigma}\,.
\end{equation}

By the approach in \cite{Belin:2013dva}, the inverse Laplace transform of the LHS of \eqref{eq:Laplace} can be calculated by writing it as an infinite sum. Here we give more analytic details together with our explicit expression of $S_n$. We expand the LHS of \eqref{eq:Laplace} as
\begin{align}
    \exp [(1-n)S_n]&=\exp (s_0-s_1-s_0n)\exp \biggl(\sum\limits_{i=1}^{\infty}{u_i n^{-i}}\biggr)\nonumber\\
    &=\biggl(1+\sum\limits_{i=1}^{\infty}{v_i n^{-i}}\biggr) \exp (s_0-s_1-s_0n),
  \label{eq:renyi-expand-2}
\end{align}
where $u_i$ and $v_i$ can be written as the function of expansion coefficients of $S_n$, i.e., 
\begin{equation}
  u_i=s_i-s_{i+1},\qquad v_{i}=\frac{1}{i!}B_{i}(u_1,\, 2!u_2,\, \ldots,\, i! u_i),
  \label{eq:renyi-expand-2-coe}
\end{equation}
where $B_i$ are the complete exponential Bell polynomials, which can be expressed as an infinite sum (see appendix~\ref{sec:math-notes} for details).

The inverse Laplace transform of~\eqref{eq:renyi-expand-2} is given by the convolution theorem
\begin{align}
    \rho(\lambda)&=\frac{\exp(s_0-s_1)}{\lambda} \biggl(\lambda_1 \delta(\lambda_1-\lambda)+\sum_{i=0}^{\infty}{\frac{v_{i+1}}{i!} \bigl(\ln(\lambda_1/\lambda)\bigr)^{i}}\theta(\lambda_1-\lambda)\biggr)\nonumber\\
    &=\exp(s_0-s_1)\biggl(\delta(\lambda_1-\lambda)+\frac{\theta(\lambda_1-\lambda)}{\lambda}\sum_{i=0}^{\infty}{\frac{v_{i+1}}{i!} \bigl(\ln(\lambda_1/\lambda)\bigr)^{i}}\biggr).
    \label{eq:ads4-spectrum-gene}
\end{align}
Similar to the special case \eqref{eq:spectrum-sqrt3}, the general spectrum also includes a Dirac delta function and Heaviside step function determined by the largest eigenvalue $\lambda_1$, which provides an example to verify the analysis above that the spectrum consists of one discrete eigenvalue and the continuous part. The largest eigenvalue of the spectrum is given by the constant term of the R\'{e}nyi entropies
\begin{align}	
    &\lambda_1=\exp(-s_0)\nonumber\\
    &=	
    \begin{cases}	
      \exp \Bigl[\frac{L^2}{4G}\bigl(2\lvert 1-\alpha ^2 \rvert \cdot \lvert 1-3\alpha^2\rvert ^{\frac{1-3\alpha ^2}{2 (1+\alpha ^2)}} \lvert 3-\alpha ^2\rvert ^{\frac{\alpha ^2-3}{2 (1+\alpha ^2)}}-1\bigr)V_{\Sigma}\Bigr],&0\leq \alpha \leq 1/\sqrt{3} \text{ or } \alpha \geq \sqrt{3},\\	
      \exp \bigl[\frac{L^2}{36 G}(2\sqrt{3}-9)V_{\Sigma} \bigr],&1/\sqrt{3}<\alpha<\sqrt{3}.	
    \end{cases}	
\end{align}
By comparing \eqref{eq:ads4-spectrum-gene} with the ansatz \eqref{eq:renyi-eigen}, the degeneracy of the eigenvalue $\lambda_1$ is
\begin{equation}	
  d_1=\exp (s_0-s_1)=	
  \begin{cases}	
    \exp \Bigl[\frac{L^2}{4G} \Bigl(\frac{ 1-3 \alpha ^2}{3-\alpha ^2} \Bigr)^{\frac{1-\alpha ^2}{1+\alpha ^2}} V_{\Sigma}\Bigr],&0\leq \alpha \leq 1/\sqrt{3}\ \text{or}\ \alpha \geq \sqrt{3},\\	
    \exp \bigl(\frac{L^2}{12G}V_{\Sigma}\bigr), &1/\sqrt{3}<\alpha<\sqrt{3},	
  \end{cases}	
\end{equation} 
which reduces to $d_1=1$ when $\alpha=1/\sqrt{3}$ and $\sqrt{3}$. The above calculations show that the largest eigenvalue of the entanglement spectrum decreases while its degeneracy increases if we enlarge the entangling sphere. Besides, the density of the continuous part $ \bar{\rho}(\lambda)$ increases. The normalization of the spectrum~\eqref{eq:ads4-spectrum-gene} can be verified numerically to satisfy
\begin{equation}
  d_1 \lambda_1+\int_{0}^{\lambda_1}{\lambda \bar{\rho}(\lambda)d\lambda}=1.
\end{equation}

\begin{figure}
  \centering
  \includegraphics[scale=0.7]{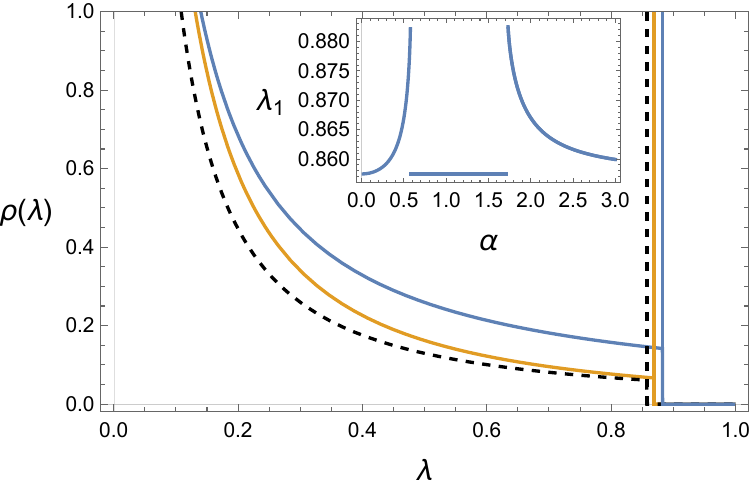}
  \caption{The entanglement spectrum as a function of $\lambda$. The dashed line is for the SAdS$_4$ black hole, while the solid lines are for hairy black holes, with $\alpha=0.52$ (orange line) and $1/\sqrt{3}$ or $\sqrt{3}$ (blue line). The largest eigenvalues are 0.857, 0.869, 0.882, respectively. The largest eigenvalue as  a function of $\alpha$ is shown in the subfigure.}
  \label{fig:ads4-spectrum}
\end{figure}

\begin{figure}
  \centering
  \includegraphics[scale=0.7]{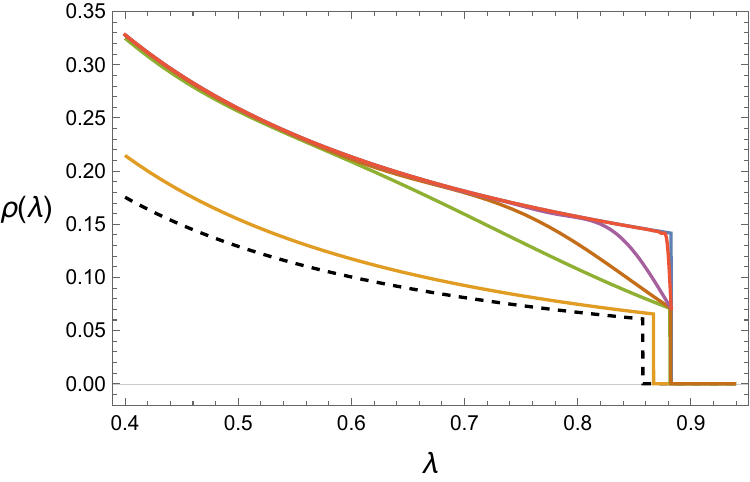}
  \caption{The entanglement spectrum near $\alpha=1/\sqrt{3}$. The dashed line is for the SAdS$_4$ black hole, while the solid lines are for hairy black holes, where the yellow, green, brown, purple, red, and blue lines are for $\alpha=0.5,\ 0.576,\ 0.577,\ 0.5773,\ 0.57735$, and $1/\sqrt{3}$, respectively. Here we plot only the continuous spectrum. }
  \label{fig:ads4-spectrum-num}
\end{figure}

As a special example, when $\alpha=1/\sqrt{3}$ or $\sqrt{3}$, we have $s_0=s_1$ and $s_i=0$ ($i>1$). Inserting them into~\eqref{eq:renyi-expand-2-coe} and~\eqref{eq:ads4-spectrum-gene}, we obtain the spectrum  
\begin{align}
    \rho(\lambda)&=\delta(\lambda_1-\lambda)+\frac{1}{\lambda}\sum_{i=0}^{\infty}{\frac{s_0^{i+1}}{i!(i+1)!} \bigl(\ln(\lambda_1/\lambda)\bigr)^{i}}\theta(\lambda_1-\lambda)\nonumber\\
    &=\delta(\lambda_1-\lambda)+\frac{s_0 \theta(\lambda_1-\lambda)}{\lambda \sqrt{s_0 \ln(\lambda_1/\lambda)}}I_1\bigl(2\sqrt{s_0 \ln(\lambda_1/\lambda)}\bigr),
    \label{eq:spectrum-ads4-analy}
\end{align}
which exactly gives the result~\eqref{eq:spectrum-sqrt3}. For general cases, the entanglement spectrum is shown in figures~\ref{fig:ads4-spectrum} and \ref{fig:ads4-spectrum-num}. The spectrum in AdS$_5$ is the same as~\eqref{eq:spectrum-ads4-analy} when $\alpha=2/\sqrt{6}$ or $4/\sqrt{6}$.

We note that the expansion of the R\'{e}nyi entropies at $n=\infty$ \eqref{eq:renyi-expand-1} has a radius of convergence. From \eqref{eq:ads4-renyi-general} and \eqref{eq:ads4-renyi-co}, the expansion converges for $n$ larger than a special value $n_c$ determined by $\alpha$ ($0\leq \alpha \leq 1/\sqrt{3}$ and $\alpha \geq \sqrt{3}$),
\begin{equation}
  n_c=\frac{1+\alpha^2}{\sqrt{(3-\alpha ^2) (1-3 \alpha ^2)}},
  \label{eq:n-vilid}
\end{equation}
which increases as $\alpha$ approaches $1/\sqrt{3}$ or $\sqrt{3}$. When $\alpha \leq \sqrt{2}-1$ or $\alpha \geq \sqrt{2}+1$, the expansion can converge at $n=1$. Therefore, we shall not expect the expansion \eqref{eq:renyi-expand-1} to describe the R\'{e}nyi entropies of all orders, especially when $\alpha$ is extremely close to $1/\sqrt{3}$ or $\sqrt{3}$. This leads to non-commutativity of the $\alpha \to 1/\sqrt{3}$ and $n \to \infty$ limits of the R\'{e}nyi entropies. As long as $\alpha$ is not too close to $1/\sqrt{3}$ or $\sqrt{3}$, the spectrum given by the series \eqref{eq:ads4-spectrum-gene} is expected to explicitly describe the low-lying part (larger $\lambda$), which is determined by the large $n$ behavior of the R\'{e}nyi entropies. The spectrum for $\lambda$ close to zero may be inconsistent with \eqref{eq:ads4-spectrum-gene}, and needs to be treated differently.

For $\lambda \to 0$, the spectrum can be approximated by the saddle point method. The saddle point $n_0$ is given by 
\begin{equation}
  \frac{\partial}{\partial n}\bigl((n-1)S_n\bigr) \Bigl|_{n_0}+\ln{\lambda}=0,
  \label{eq:saddle-point}
\end{equation}
and the integral \eqref{eq:inv-laplace} is approximated as
\begin{equation}	
  \rho(\lambda \to 0)\sim \frac{1}{\lambda^{n+1}} e^{(1-n)S_{n}} \Bigl[ 2\pi \frac{\partial^2}{\partial n^2} \bigl((1-n)S_n \bigr)\Bigr]^{-1/2}\Big|_{n_0}.	
\end{equation}
For instance, the saddle point is $n_0=\sqrt{s_0/\ln (\lambda_1/\lambda)}$ when $\alpha=1/\sqrt{3}$. The consistency is convenient to verify by changing the variable $t=-\ln \lambda$
\begin{equation}
  \rho(t \to \infty) \sim \frac{1}{16\sqrt{\pi}}\bigl[s_0(t-s_0)\bigr]^{-3/4}e^ {2\sqrt{s_0(t-s_0)}},
\end{equation}
which gives the leading term of the asymptotic expansion of~\eqref{eq:spectrum-ads4-analy} at $t= \infty$. Considering the phase transition at $n=1$, all these spectra should have the same $\lambda \to 0$ behavior (as the SAdS solution), whose saddle point is $n_0= (2/t)^{1/3}/3 +\mathcal{O}(t^{-4/3})$.

As a comparison, we also employ numerical methods\footnote{Some details of numerical methods are given in \cite{Cheng:1998IO} (or see appendix~\ref{sec:math-notes}). Different methods can give different results, some of which are unreliable in our cases. Because of the great change when $\alpha$ approaches $1/\sqrt{3}$, a singularity of \eqref{eq:ads4-renyi-general}, we test the accuracy of numerical methods by comparing the cases when $\alpha$ is extremely close to $1/\sqrt{3}$ with \eqref{eq:spectrum-sqrt3}, shown in figure~\ref{fig:ads4-spectrum-num}.} to evaluate the inverse Laplace transform, which shows that the spectrum obtained from \eqref{eq:ads4-spectrum-gene} is accurate for a wide range of $\lambda$ as long as $\alpha$ is not extremely close to $1/\sqrt{3}$. Therefore, the series \eqref{eq:ads4-spectrum-gene} is sufficient to describe the low-lying entanglement spectrum, except when $\alpha$ is extremely close to $1/\sqrt{3}$ (or $\sqrt{3}$). The low-lying spectrum for $\alpha$ close to $1/\sqrt{3}$ is plotted in figure~\ref{fig:ads4-spectrum-num} numerically, from which we can see the spectrum has unique behavior when $\lambda$ approaches $\lambda_1$:
\begin{itemize}
  \item When $\alpha$ is not extremely close to $1/\sqrt{3}$, the spectrum gradually changes with respect to $\alpha$ as a whole, and the largest eigenvalue increases, as expected. 
  \item When $\alpha$ is extremely close to $1/\sqrt{3}$, the continuous part of the spectrum near $\lambda_1$ experiences dramatical change: the spectral density of large eigenvalues increases with $\alpha$, while the spectrum with small eigenvalues is almost fixed. The degeneracy of $\lambda_1$ decreases while the location keeps almost fixed.
\end{itemize}

In the procedure from \eqref{eq:renyi-expand-1} to \eqref{eq:ads4-spectrum-gene}, we neglect the third-order phase transition of the R\'{e}nyi entropies because the spectrum of the low-lying part is not affected by the phase transition at $n=1$, based on the following observations. First, the spectrum near the largest eigenvalue $\lambda_1$ is entirely determined by the large $n$ behavior of the R\'{e}nyi entropies. Second, the saddle point approximation shows that the spectrum with $\lambda \to 0$ is dominated by the R\'{e}nyi entropies with $n \to 0$. Furthermore, when we perform the expansion~\eqref{eq:renyi-expand-1} and \eqref{eq:renyi-expand-2}, we find that the number of terms and the precision of the expansion would significantly affect the R\'{e}nyi entropies \eqref{eq:renyi-expand-1} beyond the convergent region. However, the spectrum obtained from these expansions does not show any difference for a wide range of $\lambda$. Another similar observation comes from the comparison between some numerical methods and~\eqref{eq:ads4-spectrum-gene}, which shows consistency until $\lambda$ is small enough (near zero). These observations suggest that the spectrum of different $\lambda$ is affected by different parts of the R\'{e}nyi entropies: the spectrum with large $\lambda$ is not significantly affected by small $n$. As a consequence, the phase transition at $n=1$ is merely embodied in the spectrum with $\lambda$ close to zero, and the spectra for different $\alpha$ have the same asymptotic behavior (as the SAdS black hole) at $\lambda \to 0$.

Besides the phase transition at $n=1$ where $\partial_n^2\bar{S}_n$ is discontinuous, a zeroth-order phase transition exists for $1/\sqrt{3}<\alpha<\sqrt{3}$ ($\alpha\neq 1)$ at $n=n_\text{m}$ where $\bar{S}_n$ is discontinuous; see figure~\ref{fig:ads4-renyi}. The approximation~\eqref{eq:saddle-point} implies another saddle point $n_{\text{m}}$. So the $\lambda \to 0$ behavior of the entanglement spectrum may be different from the SAdS one, although they have the same $n \to 0$ behavior of the R\'{e}nyi entropies. However, due to the same large $n$ behavior of the R\'{e}nyi entropies, the spectrum near $\lambda_1$ should be the same as the SAdS case if $\alpha$ is not extremely close to $1/\sqrt{3}$ or $\sqrt{3}$. It would be interesting to see how the phase transitions affect the full entanglement spectrum more precisely.

\section{Summary and discussion}
In this paper, we have calculated the R\'{e}nyi entropies (with a spherical entangling surface) and entanglement spectrum from a class of hyperbolic black holes with scalar hair in terms of the conformal mapping approach \cite{Casini:2011kv,Hung:2011nu}. The main conclusions are as follows:

\begin{itemize}
\item By employing a class of hyperbolic black holes with scalar hair as a one-parameter family generalization of the MTZ black hole, we explicitly obtain the holographic R\'{e}nyi entropies. The $n \to 1$ limit of these R\'{e}nyi entropies gives the same entanglement entropy, while the R\'{e}nyi entropies with $n>1$ are affected by the scalar field. In special cases when $\alpha=1/\sqrt{3}$ or $\sqrt{3}$ (in AdS$_4$) and $\alpha=2/\sqrt{6}$ or $4/\sqrt{6}$ (in AdS$_5$), which correspond to special cases of STU supergravity, the R\'{e}nyi entropies have the same $n$ dependence as the universal result in 2D CFTs.

\item The zeroth-order and third-order phase transitions of black holes lead to the discontinuity of the R\'{e}nyi entropies and their second derivatives at special values $n_{\text{m}}$, which depend on the phase transition temperature of black holes.

\item From the R\'{e}nyi entropies that are analytic at $n=\infty$, we calculate the entanglement spectrum as an infinite sum by means of the exponential Bell polynomials. The result includes a Dirac delta function and a Heaviside step function determined by the large $n$ limit of the R\'{e}nyi entropies. This approach is verified to be accurate for a wide range of eigenvalues by comparing it with numerical calculations. We take a closer look at the spectrum when $\alpha$ is extremely close to $1/\sqrt{3}$ (in AdS$_4$), which shows different properties. 
\end{itemize}

The following topics need further investigation: (i)  It would be intriguing to explicitly show how phase transitions are embodied in the entanglement spectrum. (ii) This work focuses on the neutral limit, while the charged hyperbolic black holes include more complex phase structures and are worth investigating. (iii) In special cases that intersect with STU supergravity, it would be interesting to compare our work with the supersymmetric R\'{e}nyi entropies studied in \cite{Hosseini:2019and}.

\acknowledgments
We thank Yi Wang for helpful conversations. This work was supported in part by the NSF of China under Grant No. 11905298 and the Undergraduate Base Scientific Research Project under Grant No. 20211222.

\appendix

\section{Black hole thermodynamics with generalized boundary condition}
\label{sec:bh-thermo}

Since the thermodynamics of the hyperbolic black holes plays a key role in calculating the R\'{e}nyi entropies, we study the holographic renormalization of the black holes with scalar hair by giving the boundary counter terms and generalized boundary conditions.

The free energy of a black hole is calculated from the renormalized on-shell action with proper boundary counter terms. For our Einstein-scalar system in AdS$_4$ with $m^2L^2=-2$, we consider the following boundary terms\footnote{We obtain exactly the same renormalized on-shell action as in \cite{Caldarelli:2016nni,Faulkner:2010gj}, although the expressions of the boundary terms are slightly different.}
\begin{equation}
S_\partial=\frac{1}{16\pi G}\int d^3x\sqrt{-\gamma}\biggl(2K-\frac{4}{L}-LR[\gamma]-\frac{1}{2L}\phi^2-\frac{W(\phi_a)}{L\phi_a^3}\phi^3\biggr),
\label{eq:bdy-term}
\end{equation}
where $\gamma_{\mu\nu}$ is the induced metric, $K$ is the extrinsic curvature calculated from the unit normal vector pointing out of the AdS boundary, and $\phi_a$ is a coefficient in \eqref{eq:phibdy} below. The first term being the Gibbons-Hawking boundary term for a well-defined Dirichlet variational problem and the following three terms remove the divergence in the bulk action. The last term of~\eqref{eq:bdy-term} is a finite boundary term \cite{Anabalon:2015xvl}. We set the AdS radius $L=1$ in the following. By variation of the action with respect to the induced metric $\gamma_{\mu\nu}$, we obtain the boundary stress tensor
\begin{equation}
\langle T_{\mu\nu}\rangle=\frac{1}{16\pi G}\lim_{z\to 0}z^{-1}\biggl[2(Kh_{\mu\nu}-K_{\mu\nu}-2 h_{\mu\nu}+G_{\mu\nu})-\biggl(\frac{1}{2}\phi^2+\frac{W(\phi_a)}{\phi_a^3}\phi^3\biggr)h_{\mu\nu}\biggr],
\label{eq:Tmunu}
\end{equation}
where $z$ is the AdS radial coordinate in \eqref{eq:ansatz} below, and $G_{\mu\nu}$ is the Einstein tensor calculated from $\gamma_{\mu\nu}$.

We formulate the first law of thermodynamics without imposing a boundary condition, and then show that the first law of thermodynamics is unmodified by a scalar charge with a generalized boundary condition $\phi_b=W'(\phi_a)$, whose special cases correspond to multi-trace deformations in the dual CFT. We will follow the procedure in \cite{Li:2020spf} with a generalization for imposing boundary conditions of the scalar field. The choice $W(\phi_a)=c\phi_a^3$ ($c$ is a constant) gives the same result as in \cite{Li:2020spf}.

For convenience of numerical calculations, and to see the difference between planar and hyperbolic/spherical black holes, we use the metric ansatz
\begin{equation}
ds^2=\frac{1}{z^2}\biggl(-g(z)e^{-\chi(z)}dt^2+\frac{dz^2}{g(z)}+d\Sigma_{2,k}^2\biggr),\label{eq:ansatz}
\end{equation}
where $k=0,-1,1$ are for planar, hyperbolic, and spherical black holes, respectively. In practice, we can write $d\Sigma_{2,k}^2$ as
\begin{equation}
d\Sigma_{2,k}^2=\frac{dx^2}{1-kx^2}+(1-kx^2)d\varphi^2\,.
\end{equation}
The AdS boundary is at $z=0$, and the horizon is at $z=z_h$. Unlike the $k=0$ case in which $z_h$ can be fixed to be $z_h=1$ by scaling symmetries, $z_h$ is an extra parameter when $k\neq 0$. The equations of motion for the metric and the scalar field are
\begin{align}
& g'-\left(\frac{\chi'}{2}+\frac{3}{z}\right) g-\frac{1}{2z}V(\phi)+kz=0\,,\\
& \chi'-\frac{1}{2}z\phi'^2=0\,,\\
& \phi ''+\left(\frac{g'}{g}-\frac{\chi'}{2}-\frac{2}{z}\right)\phi'-\frac{1}{z^2 g}V'(\phi)=0\,.
\end{align}

Near the horizon $z=z_h$, the asymptotic behavior of the functions is
\begin{align}
g &=\bar{g}_1(z_h-z)+\bar{g}_2(z_h-z)^2+\cdots,\label{eq:ghor}\\
\chi &=\chi_h+\bar{\chi}_1(z_h-z)+\cdots,\\
\phi &=\phi_h+\bar{\phi}_1(z_h-z)+\cdots,
\end{align}
where $\chi_h$ and $\phi_h$ can be used to express other coefficients. The temperature and entropy density are given by
\begin{equation}
T=\frac{\bar{g}_1e^{-\chi_h/2}}{4\pi},\qquad s=\frac{1}{4Gz_h^2}\,.
\end{equation}

Near the AdS boundary $z=0$, the asymptotic behavior of the functions is
\begin{align}
g &= 1+\Bigl(k+\frac{1}{4}\phi_a^2\Bigr)z^2+g_3z^3+\cdots,\label{eq:fbdy}\\
e^{-\chi}g &= 1+kz^2+\Bigl(g_3-\frac{2}{3}\phi_a\phi_b\Bigr)z^3+\cdots,\label{eq:hbdy}\\
\phi &= \phi_a z+\phi_b z^2+\cdots,\label{eq:phibdy}
\end{align}
where higher-order coefficients generally depend on the specific potential $V(\phi)$. By substituting the boundary expansions into~\eqref{eq:Tmunu}, we obtain the stress tensor. Let $\langle T^\mu_{\;\,\nu}\rangle=\text{diag}(-\varepsilon,\,p,\,p,\,p)$, and we have
\begin{align}
\varepsilon &=\frac{1}{16\pi G}\bigl(-2 g_3+\phi_a\phi_b+W(\phi_a)\bigr),\label{eq:Tmu-ep}\\
p &=\frac{1}{16\pi G}\bigl(-g_3+\phi_a\phi_b-W(\phi_a)\bigr),
\label{eq:Tmu-p}
\end{align}
where $\varepsilon$ is the energy density and $p$ is the pressure density. The trace of the stress tensor is
\begin{equation}
\langle T^\mu_{\;\,\mu}\rangle =-\varepsilon+2p=\frac{1}{16\pi G}\bigl(\phi_a\phi_b-3W(\phi_a)\bigr).
\end{equation}

We adopt a general procedure developed by Wald~\cite{Wald:1993nt} to formulate the first law of black hole thermodynamics. By variation of parameters in a $(d+1)$-dimensional solution, there is a closed $(d-1)$-form $\delta Q-i_\xi\Theta$ with $\xi$ being a Killing vector. Here we take $\xi=\partial/\partial_t$. Thus, the variation of a Hamiltonian is defined by the integral
\begin{equation}
\delta H=\int_{\Sigma^{(d-1)}}(\delta Q-i_\xi\Theta)\,,
\end{equation}
where $\Sigma^{(d-1)}$ is a $(d-1)$-dimensional surface at constant $t$ and $z$.  A key observation in~\cite{Wald:1993nt} is that $\delta H$ is a radially conserved quantity, i.e., $\delta H$ is independent of $z$, and takes the same value at the horizon and the AdS boundary. For the construction of $\delta H$ in Einstein-scalar systems, see~\cite{Liu:2013gja,Lu:2014maa}.

For the Einstein-scalar system~\eqref{eq:action4} with the metric~\eqref{eq:ansatz}, $\delta H$ at radius $z$ is given by 
\begin{equation}
\delta H=-\frac{V_\Sigma}{16\pi G} z^{-2}e^{-\chi/2}\biggl(\frac{2}{z}\delta g-g\phi'\delta\phi\biggr),
\end{equation}
where $V_\Sigma$ is the volume of $d \Sigma_{2,k}^2$. First we evaluate $\delta H$ at the horizon. From \eqref{eq:ghor}, we have $\delta g|_{z=z_h}=\bar{g}_1\delta z_h$, and we obtain
\begin{equation}\label{horizonH}
\frac{\delta H}{V_\Sigma}\Bigr|_{z=z_h}=-\frac{1}{8\pi G} e^{-\chi(z_h)/2}\bar{g}_1 z_h^{-3}\delta z_h= T\delta s\,,
\end{equation}
where we have used $\delta s=\delta(1/(4Gz_h^2))=-1/(2Gz_h^3)\delta z_h$. Evaluating $\delta H$ at the AdS boundary gives
\begin{equation}
\frac{\delta H}{V_\Sigma}\Bigr|_{z=0}=-\frac{1}{16\pi G}\lim_{z\to 0} z^{-2} e^{-\chi/2}\biggl(\frac{2}{z}\delta g-g\phi'\delta\phi\biggr)=\frac{1}{16\pi G}(-2\delta g_3+\phi_a\delta\phi_b+2\phi_b\delta\phi_a)\,.
\end{equation}
The variation of the energy density~\eqref{eq:Tmu-ep} gives
\begin{equation}
\delta\varepsilon=\frac{1}{16\pi G}\bigl(-2\delta g_3+\phi_a\delta\phi_b+(\phi_b+W'(\phi_a))\delta\phi_a\bigr).
\end{equation}
Then we obtain
\begin{equation}
\frac{\delta H}{V_\Sigma}\Bigr|_{z=0}=\delta\varepsilon+\frac{1}{16\pi G}(\phi_b- W'(\phi_a))\delta\phi_a\,.
\end{equation}
We will set $16\pi G=1$ for simplicity. By equating $\delta H$ at the horizon and the boundary, we obtain the first law of thermodynamics:
\begin{equation}
\delta\varepsilon=T\delta s-(\phi_b-W'(\phi_a))\delta\phi_a\,.
\end{equation}
We can clearly see that upon imposing a generalized boundary condition $\phi_b=W'(\phi_a)$, the first law of thermodynamics is unmodified by the scalar field. 

In other words, for a boundary condition parametrized by $\phi_b=W'(\phi_a)$, we can always formulate the first law of thermodynamics $d\varepsilon=Tds$ (unmodified by a scalar charge) if we choose the boundary terms~\eqref{eq:bdy-term}. Special cases include multi-trace deformations of the dual CFT. The double-trace deformation corresponds to $\phi_b/\phi_a=\kappa$, and thus $W(\phi_a)=(1/2)\kappa\phi_a^2$. The triple-trace deformation corresponds to $\phi_b/\phi_a^2=\tau$, and thus $W(\phi_a)=(1/3)\tau\phi_a^3$.

The free energy is calculated by the renormalized on-shell action as $F/T=-(S+S_\partial)$, where $S$ is the bulk action and $S_\partial$ is the boundary counter terms. The free energy density is the free energy per volume: $\mathsf{f}=F/V_\Sigma$. Evaluating the bulk action gives
\begin{align}
-S &=\int d^3x\int_{z_h}^{0}dz\biggl[\biggl(\frac{2}{z^3}ge^{-\chi/2}\biggr)'+2kz^{-2}e^{-\chi/2}\biggr]\\
&=T^{-1}V_\Sigma\,\frac{2}{z^3}ge^{-\chi/2}\biggr|_{z=0}+2kT^{-1}V_\Sigma\int_{z_h}^0 z^{-2}e^{-\chi/2}dz.
\end{align}
When $k=0$, the bulk Lagrangian is a total derivative \cite{Hartnoll:2008kx}, and thus we can express the action in terms of boundary quantities. When $k\neq 0$, however, the action will inevitably involve an integral from the horizon to the boundary. Consequently, the free energy density is given by
\begin{equation}
\mathsf{f}=\underbrace{g_3-\phi_a\phi_b+W(\phi_a)}_{\small\begin{array}{c}
-p
\end{array}}+2k\biggl(\frac{1}{z}+\int_{z_h}^z \bar{z}^{-2}e^{-\chi(\bar{z})/2}d\bar{z}\biggr)\biggr|_{z=0}.
\label{eq:sff}
\end{equation}

We can verify that the thermodynamic relation $\mathsf{f}=\varepsilon-Ts$ is satisfied by using a radially conserved quantity \cite{Cai:2020wrp}
\begin{align}
    \mathcal{Q} &=z^{-2}e^{\chi/2}(ge^{-\chi})'+2k\int z^{-2}e^{-\chi/2}dz\nonumber\\
    &\equiv z^{-2}e^{\chi/2}(ge^{-\chi})'+2k\mathcal{P}(z)\,,
\label{eq:conservQ}
\end{align}
which satisfies $\mathcal{Q}'(z)=0$. Evaluating $\mathcal{Q}$ at the horizon gives
\begin{equation}
\mathcal{Q}|_{z=z_h}=-Ts+2k\mathcal{P}(z_h)\,.
\label{eq:calQ-hor}
\end{equation}
Evaluating $\mathcal{Q}$ at the AdS boundary gives
\begin{equation}
\mathcal{Q}|_{z=0}=\underbrace{3 g_3-2\phi_a\phi_b}_{\small\begin{array}{c}
-(\varepsilon+p)
\end{array}}+2k\biggl(\frac{1}{z}+\mathcal{P}(z)\biggr)\biggr|_{z=0}\,,
\label{eq:calQ-bdy}
\end{equation}
where we have used \eqref{eq:Tmu-ep} and \eqref{eq:Tmu-p}. By equating \eqref{eq:calQ-hor} and \eqref{eq:calQ-bdy}, we obtain the free energy density $\mathsf{f}=\epsilon-Ts$, which is exactly the same as~\eqref{eq:sff}. We can see that when $k=0$, we have $\mathsf{f}=-p$. For hyperbolic/spherical black holes, in contrast, the free energy depends on a bulk integral. Nevertheless, the first law of thermodynamics is valid for all $k$. We have $d\mathsf{f}=-s d T$ for the boundary condition $\phi_b=W'(\phi_a)$ together with the boundary terms~\eqref{eq:bdy-term}.

With the analytic solution in hand, we can verify the first law of thermodynamics with the boundary condition given by a triple-trace deformation. For the boundary condition given by a double-trace deformation, we can numerically solve the Einstein-scalar system~\eqref{eq:action4} with the potential~\eqref{eq:3-exp-V} by the numerical techniques described in \cite{Hartnoll:2008kx}. Integrating out from the horizon to the AdS boundary gives a map
\begin{equation}
(z_h,\,\phi_h)\mapsto (\phi_a,\,\phi_b,\,g_3)\,.
\end{equation}
Upon imposing the boundary condition $\phi_b=\kappa\phi_a$, we obtain a two-parameter family of solutions, where the two parameters are two dimensionless combinations of ($z_h$, $T$, $\kappa$). By extracting the thermodynamic quantities $\varepsilon$, $T$, and $s$ from the numerical solution, we have verified that the first law of thermodynamics $d\varepsilon=Tds$ is satisfied.

\section{Mathematical notes}
\label{sec:math-notes}

\subsection{Inverse Laplace transform}
The Laplace transform of a real variable function $f(t)$ is defined as
\begin{equation}
  F(s)=\mathcal{L}[f(t)](s)=\int_{0}^{\infty}{f(t)e^{-st}dt}.
\end{equation}
The inverse Laplace transform can be obtained by Mellin's inversion formula
\begin{equation}
  f(t)=\mathcal{L}^{-1}[F(s)](t)=\frac{1}{2\pi i}\lim_{T \to \infty} \int_{\gamma-iT}^{\gamma+iT}{F(s)e^{st}dt}=\sum_{i=1}{\text{Res}[e^{st}F(s),s_i]},
  \label{eq:inv-laplace-def}
\end{equation}
where the integral is taken over a vertical line with $\text{Re}(s)=\gamma$, and $\gamma$ is a real number ensuring no singularities on the right side of this line. Thus the result can be obtained by considering the residues of $e^{st}F(s)$ at all isolated singularities $s_i$ on the complex plane. Two basic inversions related to our calculations are 
\begin{equation}
  \mathcal{L}^{-1}[e^{-\tau s}](t)=\delta(t-\tau),\qquad \mathcal{L}^{-1}\Bigl[\frac{e^{-\tau s}}{s}\Bigr](t)=\theta(t-\tau).
\end{equation}

Generally, the Laplace transform is not easy to invert, and thus we would apply numerical methods to evaluate the integral \eqref{eq:inv-laplace-def}. The accuracy and efficiency of different methods were tested and compared in terms of many analytic results \cite{Cheng:1998IO,Davies:19791}. We make our choice by comparing the spectrum (with $\alpha \to 1/\sqrt{3}$) obtained from direct numerical calculations with \eqref{eq:spectrum-sqrt3}. It turns out that the Stehfest method \cite{Stehfest:1970Al} and the Piessens Gaussian Quadrature (PGQ) method \cite{Piessens:1971Ga} behave better in our cases. The Stehfest method approximates the inversion by the following sum 
\begin{equation}
  f(t)\approx \frac{\ln 2}{t} \sum_{n=1}^{N}c_n F\Bigl(\frac{n \ln 2 }{t}\Bigr),
\end{equation}
with
\begin{equation}
  c_n=(-1)^{n+N/2}\sum_{k=(n+1)/2}^{\min \{n,N/2\}}{\frac{k^{N/2}(2k)!}{(N/2-k)!k!(k-1)!(n-k)!(2k-n)!}},
\end{equation}
where $N$ is an even number. The PGQ method is to evaluate the sum 
\begin{equation}
  f(t)\approx \frac{1}{t}\sum_{i=1}^{N}{w_i x_i F\Bigl(\frac{x_i}{t}\Bigr)},
\end{equation}
where $x_i$ are zero points of an orthogonal polynomial $P_i(x)$ with the recursive relation
\begin{equation}
  \begin{split}
    P_0(x)&=1,\qquad P_1(x)=x-1,\\
    P_i(x)&=\Bigl[2(2i-1)x+\frac{2}{2i-3}\Bigr] P_{i-1}(x)+\frac{2i-1}{2i-3}P_{i-2}(x).
  \end{split}
\end{equation}
And the weight is 
\begin{equation}
  w_i=(-1)^{N-1} \frac{1}{N } \biggl(\frac{2N-1}{x_i P_{N-1}(1/x_i)}\biggr)^2.
\end{equation}
We find that the PGQ method (figure~\ref{fig:ads4-spectrum-num}) is more accurate when $\lambda$ is close to the largest eigenvalue $\lambda_1$, while the Stehfest method is accurate when $\lambda$ is not so close to $\lambda_1$.

\subsection{The Bell polynomials}
In \eqref{eq:renyi-expand-2-coe}, we use the complete exponential Bell polynomials to express the expansion coefficients. The complete Bell polynomials have the generating function
\begin{equation}
  B_0=1,\qquad B_i(x_1,\ldots,x_i)=\biggl(\frac{\partial}{\partial t}\biggr)^i \exp{\biggl(\sum_{j=1}^{i}x_j \frac{t_j}{j!}\biggr)\bigg|_{t=0}}.
\end{equation}
These polynomials can also be expressed as an infinite sum of the partial exponential Bell polynomials
\begin{equation}
  B_i(x_1,\ldots,x_i)=\sum_{j=1}^{i}{B_{i,j}(x_1,\ldots,x_{i-j+1})},
\end{equation}
with
\begin{equation}
  B_{i,j}(x_1,\ldots,x_{i-j+1})=\sum_{\{k\}}^{}{\frac{i!}{k_1! \cdots k_{i-j+1}!}\prod_{l=1}^{i-j+1}{\Bigl(\frac{x_l}{l!}\Bigr)^{k_l}}},
\end{equation}
where $\{k\}$ means that the sum is taken over all sequences $\{k_1 \sim k_{i-j+1}\}$ that satisfy
\begin{equation}
  \sum_{l=1}^{i-j+1}{k_l}=j, \qquad \sum_{l=1}^{i-j+1}{l\,k_l}=i.
\end{equation}

\end{document}